Master Thesis - Master of Computer Science# A Scalable, Linear-Time Dynamic Cutoff Algorithm for Molecular Simulations of Interfacial Systems

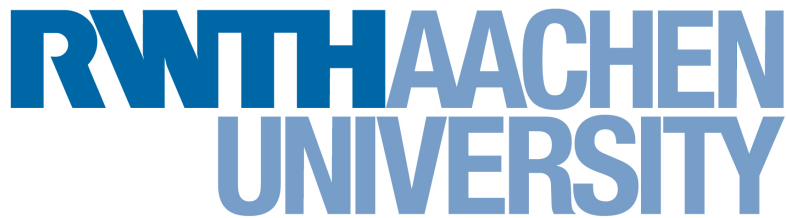

Paul Springer

RWTH Aachen University

Paul.Springer@RWTH-Aachen.de

Supervised by

Prof. Paolo Bientinesi[1], Ph.D.   &   Prof. Ahmed E. Ismail[1], Ph.D.

[1]Institute for Advanced Study in Computational Engineering Science

## Declaration of Independence

I hereby declare that I am the sole author of this thesis and that I only used the resources listed in the reference section.

Aachen, Germany
Mai 31, 2014

Paul Springer

# Abstract

This master thesis introduces the idea of dynamic cutoffs in molecular dynamics simulations, based on the distance between particles and the interface, and presents a solution for detecting interfaces in real-time. Our dynamic cutoff method (DCM) exhibits a linear-time complexity as well as nearly ideal weak and strong scaling. The DCM is tailored for massively parallel architectures and for large interfacial systems with millions of particles. We implemented the DCM as part of the LAMMPS open-source molecular dynamics package and demonstrate the nearly ideal weak- and strong-scaling behavior of this method on an IBM BlueGene/Q supercomputer. Our results for a liquid/vapor system consisting of Lennard-Jones particles show that the accuracy of DCM is comparable to that of the traditional particle-particle particle-mesh (PPPM) algorithm. The performance comparison indicates that DCM is preferable for large systems due to the limited scaling of FFTs within the PPPM algorithm. Moreover, the DCM requires the interface to be identified every other MD timestep. As a consequence, this thesis also presents an interface detection method which is (1) applicable in real time; (2) parallelizable; and (3) scales linearly with respect to the number of particles.

## Acknowledgments

I like to express my deep gratitude to Prof. Paolo Bientinesi and Prof. Ahmed E. Ismail for their superb support throughout this thesis. Only their continued confidence in my work and their guidance made this thesis possible. A special thanks goes to Prof. Benjamin Berkels who added the missing *image-segmentation* knowledge and greatly improved the quality of this work. Moreover, I thank Dr. Edoardo Di Napoli for granting me access to the JUQUEEN supercomputer at Forschungszentrum Jülich. Last but not least I am grateful for the valuable and enlightening discussions with Dipl.-Ing. Rolf Isele-Holder and Dipl.-Ing. Daniel Tameling.

# Contents

## 1. Motivation and Contribution

Molecular dynamics (MD) simulations are ubiquitous in computational chemistry, materials science and biophysics. Their ability to study the dynamics of systems with an every growing number of particles is fueled by the exponentially increasing compute power in modern computer architectures [1]. The last decade has seen a paradigm shift from single-core CPUs to multi-core CPUs. This paradigm shift makes good scalability an absolute necessity if we want to continue to push the limits of molecular dynamics simulations to larger systems and longer time scales. Hence, we developed the dynamic cutoff method (DCM) which is especially tailored for large MD simulations that require massively parallel supercomputers to solve them.

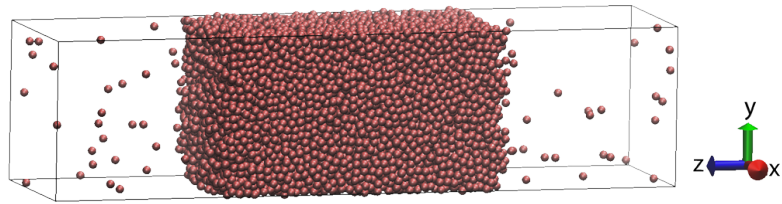

Figure 1: Interfacial system with 16000 particles. Particles in the high-density region (center) belong to the liquid phase while particles in the low-density region (left and right) belong to the vapor phase. The transition from the liquid phase to the vapor phase describes the interfaces.

Molecular dynamics simulations can be roughly broken down into short-range and long-range calculations. While long-range algorithms deal with force contributions from particles which are far apart from each other, short-range algorithms restrict themselves to particle-particle interactions within a certain distance, known as the *cutoff*. Despite the weak $r^{-n}$ scaling of short-range potentials (see Section 2 for further details) with the particle-particle distance $r$ and $n$ being larger than the dimension of interest (i.e., larger than 3 in most cases), it is not sufficient to use a classical short-range algorithm for molecular systems with interfaces [2–5]. An example of an interfacial system is illustrated in Figure 1.

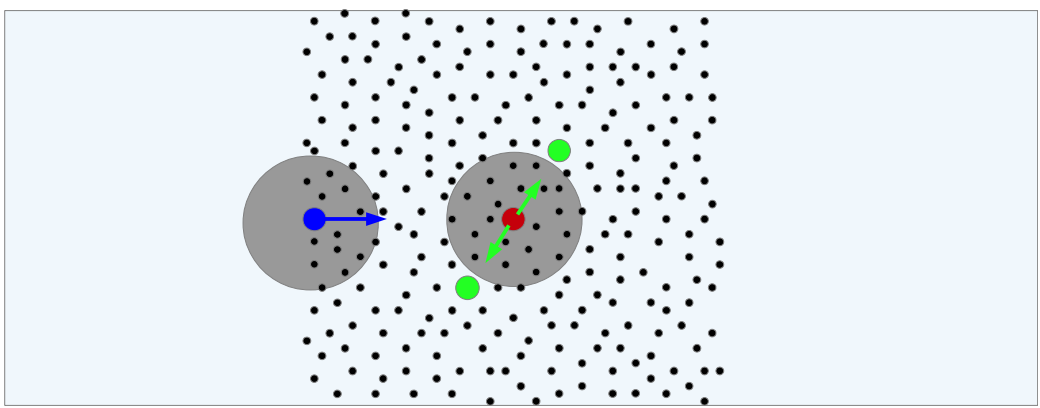

Figure 2: Interfacial system using a static cutoff. Gray area denotes the cutoff. The arrows represent the force acting on a particle.

Figure 2 illustrates an interfacial system using a static cutoff. The problem with this approach becomes apparent when looking at the red and blue particles. On the one hand it is reasonable to use a short-range method which neglects all the particle-particle interaction beyond the cutoff for the red particle for two reasons: First, as the distance $r_{i,j}$ between particles $i$ and $j$ increases the short-range forces converge to zero rather quickly (see Section 2 for further details). Second, particles within the bulk phase are arranged more or less homogeneously. As an example, accounting for the forces of the green particles on the red particle (green arrows) would not increase the accuracy because these forces have the same absolute value and point in opposite directions and thus cancel each other's contribution. For particles close to the interface (e.g., the blue particle), on the other hand, it is no longer reasonable to assume a homogeneous distribution of particles. As a result, using a cutoff method introduces a non negligible error because *interfacial* particles are missing the attractive forces from particles outside of the cutoff which is not canceled by particles on the opposite side of the cutoff (simply because there are none).

Figure 3 shows the particle density along the *z*-axis[1] for the system shown in Figure 1 with respect to three different cutoffs. Due to the neglected attractive forces of cutoff methods, small cutoffs tend to yield lower (i.e., less accurate) densities (see Figure 3) than larger cutoffs. While this effect is less observable for low temperatures (see Figure 3a), the missing force contributions from particles outside of the cutoff can lead to completely wrong results for high temperatures (see Figure 3b). To be more precise, a cutoff of $3.0\sigma$ is not sufficient to preserve the liquid-vapor interface and yields completely wrong results.

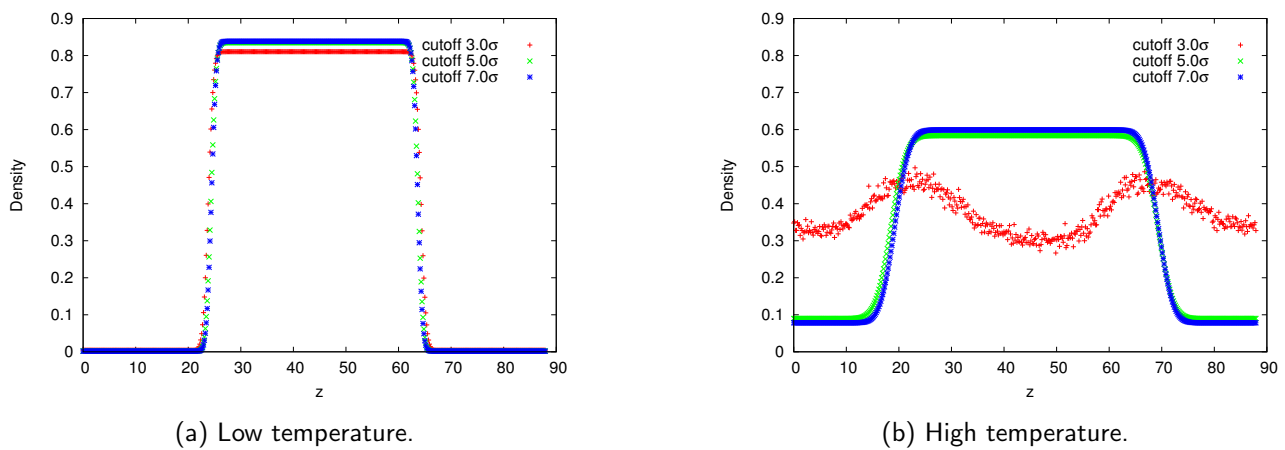

(a) Low temperature.  (b) High temperature.

Figure 3: Density profile for different cutoffs and temperatures. Note that low temperatures result in higher densities than high temperatures.

One solution to this problem is to use long-range solvers such as Ewald Sums [6–9] which scale as $\mathcal{O}(N^{3/2})$, where $N$ is the number of particles, hierarchical methods [10] and mesh-based Ewald methods [11–15], which scale as $\mathcal{O}(N \log N)$, or fast-multipole methods [16] as well as the multi-level summation methods [17, 18], which scale as $\mathcal{O}(N)$.

Despite the good asymptotic complexity of some of these solutions, they all share a common disadvantage: non-local communication. This disadvantage becomes progressively more severe as the performance gap between floating-point and memory operations keeps widening. Moreover, as the trend towards massively parallel architectures continues, the next generation of supercomputers will be much more affected by communication overhead than today's systems.

With these insights in mind, we developed the dynamic cutoff method (DCM) and — as a pre-

---

[1] All particles in the *x-y* plane are averaged.

requisite — a surface detection method within molecular simulations. DCM is designed for large interfacial systems and exhibits a strictly-local communication pattern as well as a linear-time complexity $\mathcal{O}(N)$.

The main idea of DCM is illustrated in Figure 4: We keep the computational demand to a minimum and increase the accuracy where it is required. Instead of applying a static cutoff to all particles, DCM assigns large cutoffs to particles close to an interface and small cutoffs to particles in the interior. The DCM is close in spirit to the classical (static) cutoff method [19, 20], while it exploits the fact that large cutoffs are only required for particles close to an interface.

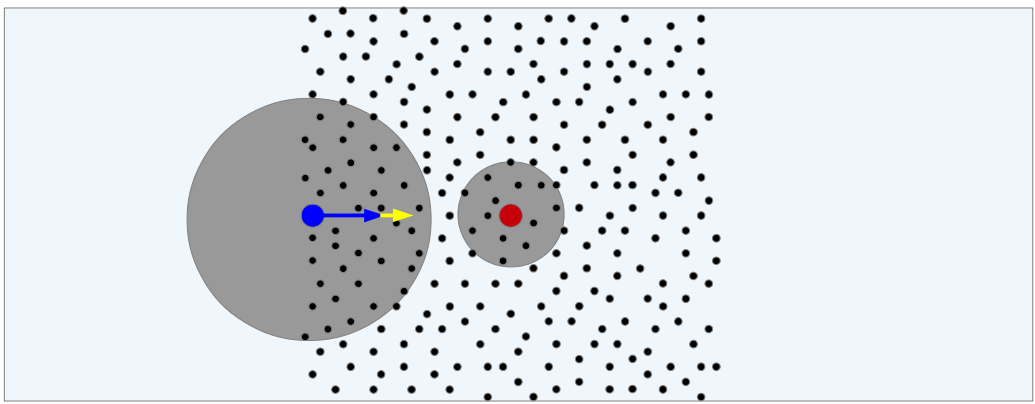

Figure 4: Interfacial system using a dynamic cutoff. Gray area denotes the cutoff. The arrows represent the force acting on a particle. The yellow arrow depicts the additional force contribution due to a larger cutoff.

**Contributions.** The key contributions of this master thesis can be summarized as follows:

- A scalable, linear-time MD algorithm for interfacial systems is introduced.

- A fair performance and accuracy comparison between DCM and PPPM, a state-of-the-art MD algorithm, is performed.

- A scalable, fast, linear-time algorithm for detecting interfaces in molecular simulations is developed. In contrast to other interface-detection methods [21–23] our algorithm is so fast that it does not affect the performance of the MD simulation.

- A cutoff-based Fast Sweeping Method (CFSM) is developed and parallelized. This algorithm greatly outperforms the standard Fast Sweeping Method [24].

The remainder of this thesis is organized as follows: Section 2 gives background information about MD simulations. Section 3 describes the details about DCM. Section 4 covers the topic of surface detection within MD. Section 5 describes the implementation details of the parallelized DCM. Section 6 compares DCM to PPPM with respect to accuracy and performance. Section 7 outlines possible directions for future work. We conclude the thesis with Section 8.

## 2. Background and Related Work

This chapter relates the work of this thesis to existing approaches and gives an overview of molecular dynamics. Moreover, we will describe short-range MD calculations in some detail and give a high-level overview of mesh-based Ewald solvers and identify their scalability bottleneck.

### 2.1. Interface Detection

The *identification of truly interfacial molecules* (ITIM) algorithm introduced by Partay et al. [21] and the *intrinsic sampling method* by Bremse et al. [22] allow to identify an interface at an atomic level. While these algorithms are restricted to planar interfaces, the generalized ITIM (GITIM) [23] is able to detect arbitrary interfaces; it is a mixture of the $\alpha$-shapes algorithm [25] and ITIM. Willard et al. introduced another surface detection algorithm for arbitrary interfaces in [26]. A comparison between different methods is presented in [27].

Our interface-detection approach is motivated by the work of Berkels et al. [28] in the field of image segmentation. As we show in this thesis, image segmentation is closely related to interface detection in molecular simulations (see Section 4). The implementation of the segmentation algorithm by Berkels et al. [28] is part of the open-source, finite-element library *QuocMesh*[2][29]. This powerful segmentation algorithm enables us to create a (1) highly scalable, (2) linear-time interface-detection method (3) which yields a good approximation to the interface within a few milliseconds. These are important properties since this algorithm will be executed every other MD timestep on thousands of CPUs. In contrast to the aforementioned algorithms, our approach is not required to represent the exact interface on an atomic level. However, we believe that the accuracy of this algorithm is quite high; a detailed accuracy analysis for the interface detection remains future work.

### 2.2. Molecular Dynamics

Molecular dynamics (MD) is a classical simulation method for many-body systems. It ranges from the simulation of the universe, stars, down to atomistic simulations of molecules and atoms. Despite the different length scales differ, the underlying idea of advancing the particles' position according to Newton's laws of motion remains the same:

$$\begin{aligned} \dot{\mathbf{x}}_i(t) &= \mathbf{v}_i(t), \\ \dot{\mathbf{v}}_i(t) &= \mathbf{a}_i(t), \end{aligned} \quad (1)$$

for $i = 1 \ldots N$ with $\mathbf{x}_i, \mathbf{v}_i, \mathbf{a}_i \in \mathbb{R}^3$ being the position, velocity and acceleration of particle $i$, respectively. Furthermore, $\mathbf{a}_i$ can be expressed as the gradient of the interaction potential $V_{ij}$:

$$\begin{aligned} \mathbf{a}_i(t) &= \frac{\mathbf{F}_i(t)}{m_i}, \quad \mathbf{F}_i(t) = -\nabla_i V(t), \\ V(t, \mathbf{x}_0, \mathbf{x}_1, \ldots, \mathbf{x}_{N-1}) &= \sum_{i,j<i} V_{ij}(r_{ij}) \end{aligned} \quad (2)$$

---

[2]QuocMesh also offers a variety of image processing algorithms based on finite differences.

where $r_{ij}$ denotes the distance between particle $i$ and particle $j$ at time $t$.

An example of an interaction potential $V_{ij}$ is the gravitational potential which is of importance for the simulation of stars:

$$V_{ij}^G(r_{ij}) = -G\frac{m_i m_j}{r_{ij}} \qquad (3)$$

In molecular simulations, on the other hand, the Lennard-Jones (LJ) potential (see Figure 5) is ubiquitous. It accounts for both the short-range repulsion of the *Pauli exclusion principle* as well as the long-range attraction of the *van der Waals forces*. Due to its fast decay with $r^{-6}$ this potential is considered to be short-ranged[3]. We use the LJ potential as the potential of choice for the remainder of this thesis. However, we want to stress that the dynamic cutoff method is also applicable to other short-range potentials as well as the short-range part of the mesh-based Ewald solvers; an extension of DCM to these potentials is left as future work.

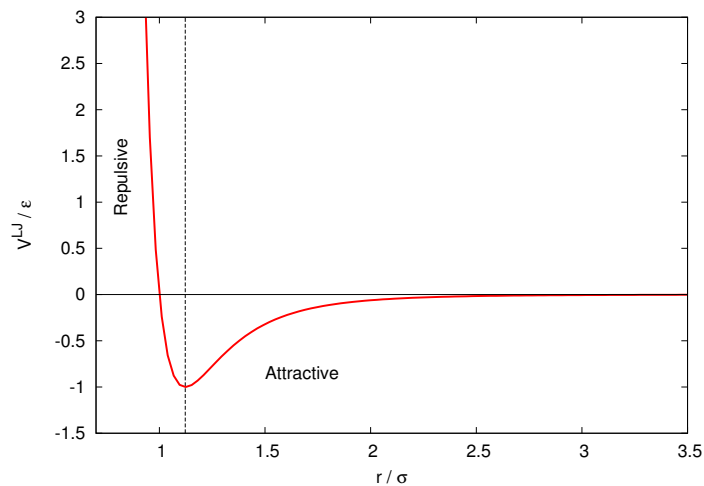

$$V^{LJ}(r_{ij}) = 4\epsilon\left(\left(\frac{\sigma}{r_{ij}}\right)^{12} - \left(\frac{\sigma}{r_{ij}}\right)^6\right) \qquad (4)$$

Figure 5: Lennard-Jones potential. (Left) Lennard-Jones potential (normalized by $\epsilon$) over the particle-particle distance $r$ (normalized by $\sigma$). $\sigma$ and $\epsilon$ being parameters of the LJ potential (see Equation 4).

To advance the position of each particle from one timestep to the next, equation 1, 2 and 4 need to be solved. The naive solution to this problem requires the interaction of each particle with every other particle, hence, resulting in a quadratic complexity $\mathcal{O}(N^2)$. Such a solution is not feasible for large systems consisting of thousands of particles. This problem becomes even more dominant as a typical MD simulation requires millions of timesteps to observe the events of interest (e.g., protein folding [30]).

Figure 6 shows a rough overview of a typical MD simulation [4]. Each MD step (i.e., one iteration of the loop) must compute the forces acting on each particle (a). Based on these forces the particle positions can be updated (b) and the process starts over. We will refine Figure 6 in the following subsections.

---

[3] All potentials which decay with $r^{-x}$ with $x$ being larger than the dimension are considered short-range potentials.
[4] For readability reasons we do not show the processing of the computed data.

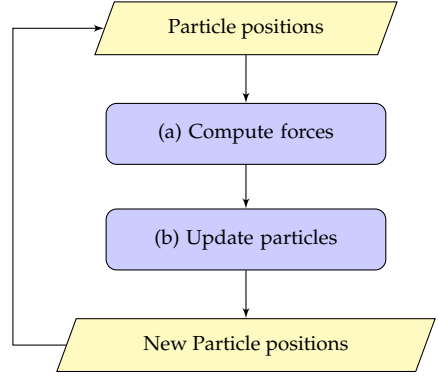

Figure 6: MD simulation overview.

### 2.2.1. Short-Range MD calculations

Due to the rapid falloff of the LJ potential with $r^{-6}$ (see Equation 4), it is reasonable to assume a maximal interaction distance $r_c$ beyond which the interaction potential is set to zero (i.e., $V(r_{ij}) = 0$, if $r_{ij} > r_c$). The truncation of the potential reduces the complexity from $\mathcal{O}(N^2)$ to $\mathcal{O}(N)$. Instead of calculating the forces for each particle $i$ with all the other particles, it is now sufficient to only interact with those particles $j$ within the cutoff $r_c$ (i.e., $r_{ij} \leq r_c$).

Hockney et al. [19] introduced the *link-cell* method, which is based on the idea of binning the particles into cells of size $r_c \times r_c \times r_c$. The binning restricts the search for all neighbors of particle $i$ to the 26 neighboring boxes and the box of particle $i$ itself. Hence, results in the desired complexity of $\mathcal{O}(N)$. The interested reader is referred to [31] for the description of a more sophisticated version of this algorithm.

Verlet et al. [20] were the first to propose the use of neighbor-lists (also referred to as verlet-lists) for each particle. The key idea behind the verlet-list is to store all particle indices $j$ which interact with particle $i$ (i.e., $r_{ij} \leq r_c$) in the neighbor-list of particle $i$; and reuse this list over multiple timesteps. To reuse this list a skin distance $r_s \in \mathbb{R}^{>0}$ is introduced and instead of only storing particle $j$ with $r_{ij} \leq r_c$ all particles with $r_{ij} \leq r_c + r_s$ are stored. The verlet-list has to be rebuild once a particle has moved at least $\frac{r_s}{2}$ away from its old position at the last neighbor-list build [32]. This technique drastically decreases the number of spurious distance calculations (i.e., distance calculations with $r_{ij} > r_c$) at line 6 of Listing 2.1. For instance, on average 84.5%[5] of the distance calculations of the cell-list method are spurious. Using the verlet-list with a skin distance $r_s = 0.1 r_c$ reduces the value to $\approx 25\%$ [6]. The choice of the skin distance $r_s$ has a strong impact on the performance: Low values of $r_s$ require a frequent rebuilding of the neighbor-list, while high values of $r_s$ would make the verlet-list less efficient. Chialvo et al. explore the dependence of $r_s$ on the cutoff, temperature and density in [32].

By combining the link-cell method with verlet-lists one bins the particles into cells and uses the obtained neighborhood information to build the verlet-list. This neighbor-list is then reused throughout the force calculation (see Listing 2.1) over multiple timesteps. The pseudocode for the neighbor-list build is outlined in Listing 2.2. We will revisit these algorithms with respect to DCM in Section 3.2.

---

[5]To account for all the force contributions within the cutoff volume $V^{cut} = 4/3\pi r_c^3$, we need to check a much bigger volume $V^{CL} = (3r_c)^3$, with $V^{cut}/V^{CL} \approx 15.5\%$.

[6]$V^{cut}/V^{verlet}$ with $V^{verlet} = 4/3\pi(r_c + r_s)^3$

**Algorithm 2.1** Force calculation
1:  **for all** atoms $i$ **do**
2:      $f_i \leftarrow 0$
3:      **for all** neighbors $k$ of $i$ **do**
4:          $j \leftarrow \text{nbrs}_i[k]$
5:          $r_{ij} \leftarrow r_i - r_j$
6:          **if** $|r_{ij}| \leq r_c$ **then**
7:              $f \leftarrow \text{forceLJ}(|r_{ij}|)$
8:              $f_i \leftarrow f_i - f \times r_{ij}$
9:              $\text{force}[j] \leftarrow \text{force}[j] + f \times r_{ij}$
10:         **end if**
11:     **end for**
12:     $\text{force}[i] \leftarrow \text{force}[i] + f_i$
13: **end for**

**Algorithm 2.2** Verlet-list build
1:  **for all** local atoms $i$ **do**
2:      $\text{nNbrs}[i] \leftarrow 0$
3:      $\text{iBin} \leftarrow \text{getBin}(i)$
4:      **for all** jBin $\in$ neighbors(iBin) **do**
5:          **for all** atoms $j$ of jBin **do**
6:              $r_{ij} \leftarrow r_i - r_j$
7:              **if** $|r_{ij}| \leq r_c$ AND $i < j$ **then**
8:                  $\text{nbrs}_i[\text{nNbrs}[i]] \leftarrow j$
9:                  $\text{nNbrs}[i] \leftarrow \text{nNbrs}[i] + 1$
10:             **end if**
11:         **end for**
12:     **end for**
13: **end for**

The refined controlflow of a MD simulation using verlet-lists is depicted in Figure 7.

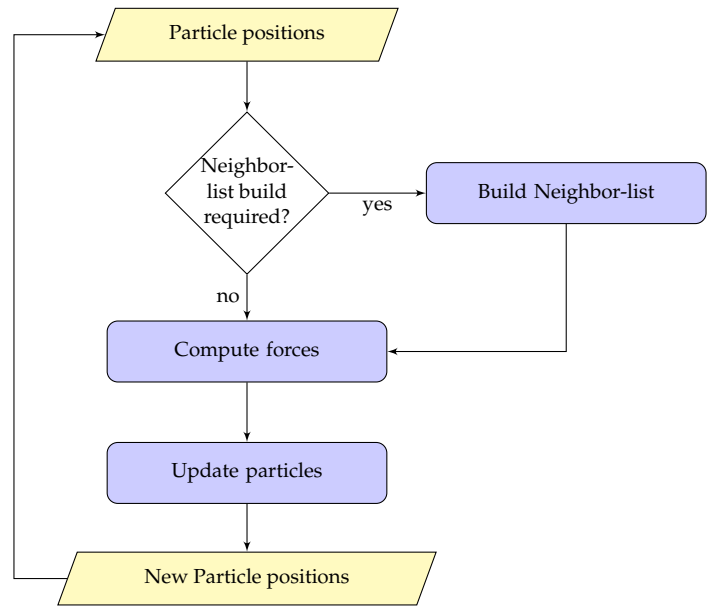

Figure 7: MD simulation overview with verlet-lists.

### 2.2.2. Adaptive Cutoff

Two weeks before submitting the thesis we received the paper of Wang et al. [33] which is an excellent work on error estimates for inhomogeneous systems of short-range force calculations. To our surprise this paper already introduced the idea of dynamic cutoffs. Despite some commonalities between their method and ours, there are also substantial differences.

Even though the idea of dynamically changing the cutoffs is the same, the approach is different. They estimate the error for each point in space and adapt the cutoff accordingly; we detect the in-

terface — the region which exhibits a large error — and compute the dynamic cutoff based on the distance to the interface. The fundamental difference between their solution and ours is that their solution requires Fast Fourier Transforms (FFTs) and therefore requires global communications and an $\mathcal{O}(N \log N)$ complexity. Our method, on the other hand, is purely based on local communication and scales linearly with the number of particles. Moreover, while their work is focused on the error estimates, our work focuses on large systems and scalability. As such we present performance results for systems with up to 39.3 million particles on up to 32,768 cores. Another noticeable difference is that their method is only applicable to systems with periodic boundary conditions while ours does not have such a constraint.

### 2.2.3. Long-Range Tail-Corrections

One approach to incorporate long-range dispersion forces to short-range calculations was introduced by Mecke et al. [34]. They showed that tail-corrections to the short-range force calculations at runtime yield a significant accuracy boost. Their approach is somewhat orthogonal to the solution we propose in this thesis. However, we strongly believe that a combination of these online tail-corrections and our solution would significantly improve the accuracy of both approaches.

### 2.2.4. Particle-Mesh Ewald Methods

The underlying idea of the Ewald summation algorithm [8] is to split the potential into two absolute convergent sums of which one is solved in real-space — using the short-range methods of the previous section — while the other sum is solved in reciprocal space.

Hockney and Eastwood [13] introduced the first mesh-based Ewald method called *particle-particle particle-mesh* (PPPM). The *particle mesh ewald* (PME) [11] and the *smooth particle mesh ewald* (SPME) [12] are further examples of mesh-based Ewald methods. Deserno et al. give a detailed comparison of these different variants [14].

In contrast to the original Ewald summation, mesh-based Ewald summations use FFTs to deal with the reciprocal sum. This reduces the complexity from $\mathcal{O}(N^{3/2})$ to $\mathcal{O}(N \log N)$. A detailed discussion of these algorithms goes well beyond the scope of this thesis; the interested reader is referred to the literature [11–14] for further details.

Figure 8a shows the bad weak-scaling behaviour of the PPPM algorithm on the BlueGene/Q system at Forschungszentrum Jülich. Despite the fact that in this experiment the number of particles per core is constant, the time to solution for PPPM (orange line) increases with the size of the systems. By contrast, our dynamic cutoff method (purple line) on the other hand shows perfect weak scaling behavior.

Figure 8b identifies the FFTs as the main cause for the limited scaling of PPPM. These findings are in line with the result of Schulz et al. [35], who also identify the FFTs as the main scalability bottleneck for the mesh-based Ewald summation. The reasons for the bad scaling of the FFTs is manifold: First, FFTs are memory-bound operations and require a lot of communication; second, FFTs exhibit an all-to-all communication pattern which prevents them from scaling perfectly; third, FFTs exhibit a $\mathcal{O}(N \log N)$ complexity and therefore scale worse than $\mathcal{O}(N)$ methods as the total number of particles $N$ increases.

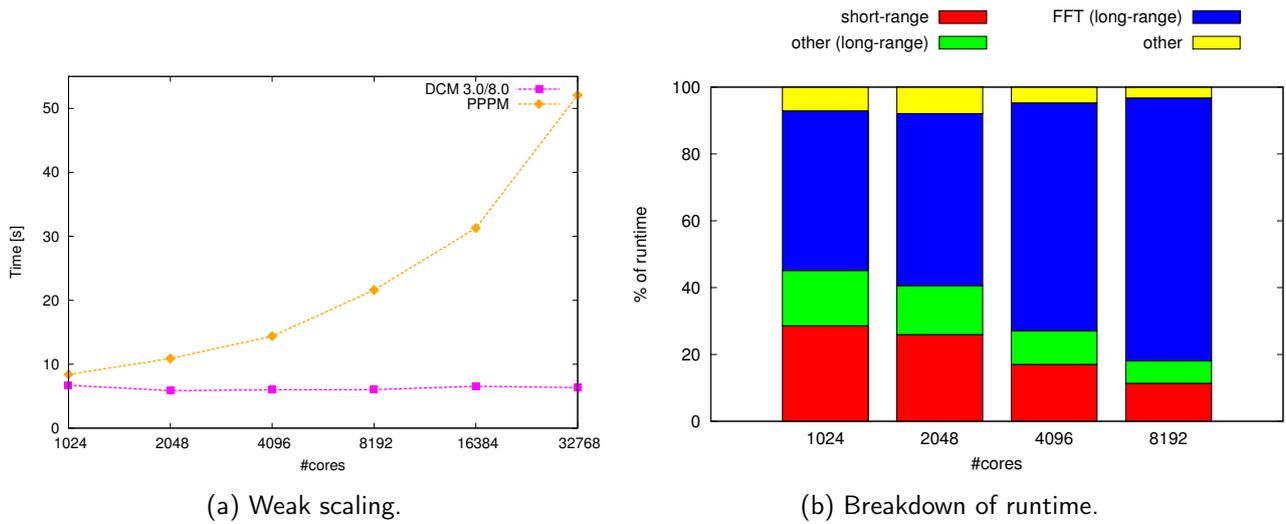

(a) Weak scaling.

(b) Breakdown of runtime.

Figure 8: (a) Weak scaling of PPPM for an interfacial system with 1200 Lennard-Jones paricles per core. (b) Percentage of PPPM spent on short-range calculations, long-range calculations and other.

The dynamic cutoff method overcomes these bottlenecks and presents an alternative solution for large interfacial systems using short-range potentials.

## 3. Dynamic Cutoff Method

The dynamic cutoff method (DCM) can be seen as an extension to the classical short-range calculations covered in Section 2.2.1. The basic idea of DCM is to change the cutoff of each particle as a function of the particle's distance to the interface. This idea is motivated by the fact that particles close to the interface require a larger cutoff than particles farther away from the interface (see Section 1). This approach both retains the low computational demands of a small cutoff while achieving the accuracy of a large cutoff (see Section 6.1).

The advantages of DCM being based on the cell-list algorithm are manifold:

- the DCM exhibits a strictly local communication pattern (i.e., neighbor-neighbor communication),
- it has a linear-time complexity with respect to the number of particles, and
- the computational demand is solely based on the number of particles and does not increase with void spaces in the domain.

However, DCM also has its disadvantages: first, the since different particles can have different cutoffs, it is possible that particle $i$ might impose a force on particle $j$ but not *vice versa* (i.e., the particle-particle interaction is no longer symmetric). This asymmetry leads to the problem that Newton's third law [7] is no longer applicable. To be precise, DCM has to store $j$ in the verlet-list of $i$ and also to store $i$ in the verlet-list of $j$ (a.k.a. *full* verlet-list) and evaluate the force between $i$ and $j$ twice. Hence, as a result increases the computational demand by a factor of two. While this sounds like a major disadvantage, we show in Section 6.2 that it is well compensated for by the better scaling of the DCM. Moreover, due to the preferred memory access patterns and fewer data dependencies, Newton's third law is also commonly neglected in GPU implementations [36, 37].

Second, load balancing becomes a challenging task because domain decomposition does not do a good job for a domain with varying numbers of particles in each subdomain. Moreover, due to the varying computational demand per particle (depending on the cutoff) an even particle distribution across the processors does not necessarily result in good load balancing either. Even though the former obstacle seems only relevant for short-range solvers, and does not have a severe impact on mesh-based Ewald solvers, it is not. Current LAMMPS [38] implementations of mesh-based Ewald solvers do not exploit the fact that subdomains with few particles would require less computations that those subdomains with many particles. Hence, mesh-based Ewald solvers would run into the same load-balancing problems once they are also able to exploit these low-density regions.

Figure 9 shows a breakdown of the DCM. The DCM, in contrast to classical short-range methods, has to compute an individual cutoff per particle — based on the particle's distance to the interface — before constructing the neighbor-list. This poses the challenge of *detecting the interface automatically*. Once the interface has been detected, it can be reused over hundreds of timesteps since the interface, in a typical MD simulation, only changes very slowly.

The following sections will focus on the computational challenges associated with the DCM and assume that an interface representation is given. A detailed discussion of the interface detection is delayed until Section 4.

---

[7] If a body $i$ exerts a force $f$ onto another body $j$, then $j$ also exerts $-f$ on $i$.

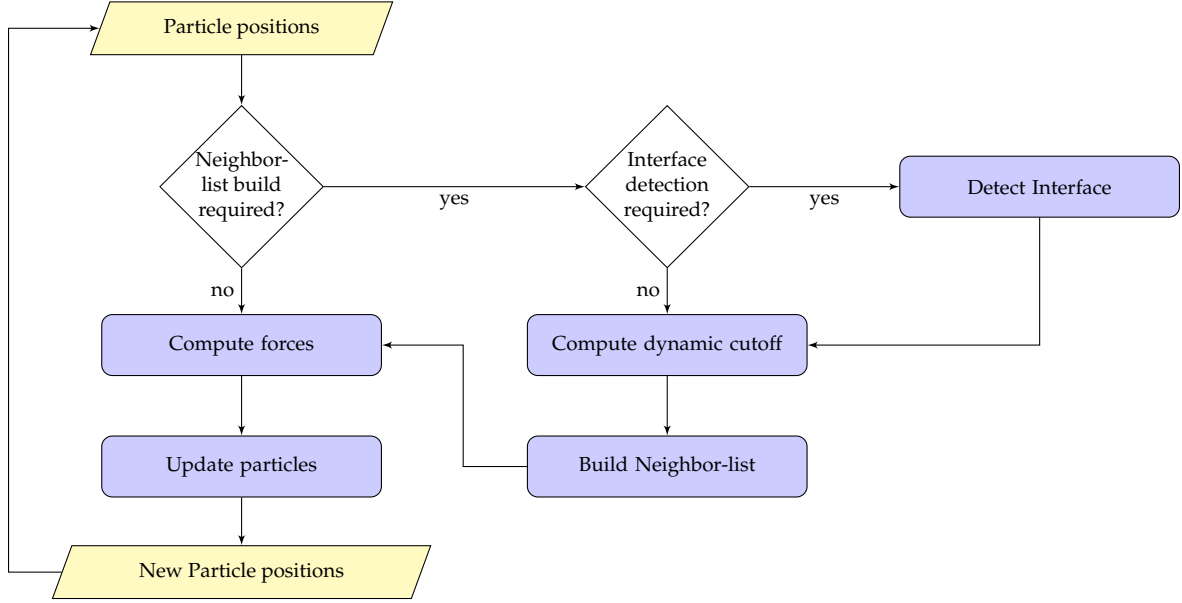

Figure 9: MD simulation overview using DCM.

The interface detection method subdivides the computational domain $L \in \mathbb{R}^3$ uniformly into a set of $N_x \times N_y \times N_z$ boxes, with $N_x, N_y, N_z \in \mathbb{N}$ respectively being the number of boxes in $x$, $y$ and $z$ direction. For now, we assume that the 3D interface is given as a set $D$ of box-interface distances $d_{i,j,k}$:

$$D = \{d_{i,j,k} \in \mathbb{R} \mid 0 \leq i < N_x, 0 \leq j < N_y, 0 \leq k < N_z\}. \tag{5}$$

Once the set of box-interface distances $D$ is available, the set of particle-interface distances $D^p$ can be computed:

$$D^p = \{d_i \in \mathbb{R} \mid 0 \leq i < N\}, \tag{6}$$

where $N$ is the number of particles. Algorithm 3.1 shows the pseudocode for a trilinear interpolation which is used to compute the approximate particle-interface distances $D^p$ from the box-interface distances $D$.

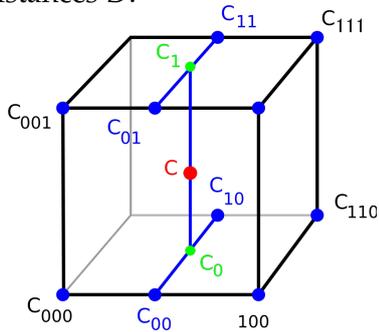

Figure 10: Trilinear interpolation. Source: [39].

**Algorithm 3.1** Verlet-list build

1: **for all** atoms $i$ **do**
2:     $(x, y, z) \leftarrow$ getLowerLeftFrontBoxIndex($i$)
3:     $C_{0,0} \leftarrow$ linearInterpolation($d_{x,y,z}, d_{x+1,y,z}, r[i]_x$)
4:     $C_{1,0} \leftarrow$ linearInterpolation($d_{x,y+1,z}, d_{x+1,y+1,z}, r[i]_x$)
5:     $C_{0,1} \leftarrow$ linearInterpolation($d_{x,y,z+1}, d_{x+1,y,z+1}, r[i]_x$)
6:     $C_{1,1} \leftarrow$ linearInterpolation($d_{x,y+1,z+1}, d_{x+1,y+1,z+1}, r[i]_x$)
7:     $C_0 \leftarrow$ linearInterpolation($C_{0,0}, C_{1,0}, r[i]_y$)
8:     $C_1 \leftarrow$ linearInterpolation($C_{0,1}, C_{1,1}, r[i]_y$)
9:     $d[i] \leftarrow$ linearInterpolation($C_0, C_1, r[i]_z$)
10: **end for**

Line 2 of Algorithm 3.1 returns the index of the lower, left, front box responsible for particle $i$. It is not sufficient to just determine the box of particle $i$ because two particles in different quadrants of

## 3 DYNAMIC CUTOFF METHOD

the same box have to interpolate over different boxes $d_{i,j,k}$ (see Figure 11). Based on the computed particle-interface distance $d_i$ in Line 9, we need to compute its dynamic cutoff $r_c[i]$. This involves the use of a *cutoff function* which will be discussed in the next subsection.

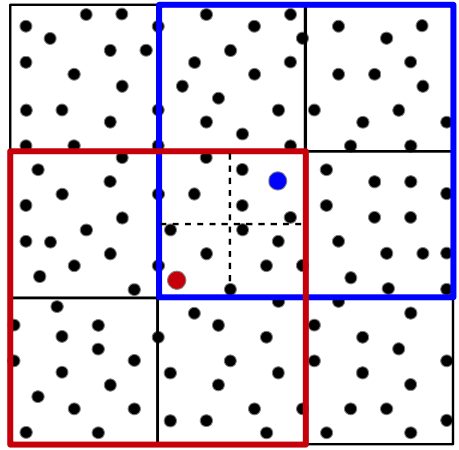

Figure 11: Two particles belonging to the same box might have to interpolate over different neighboring boxes.

### 3.1. Design of Cutoff Functions

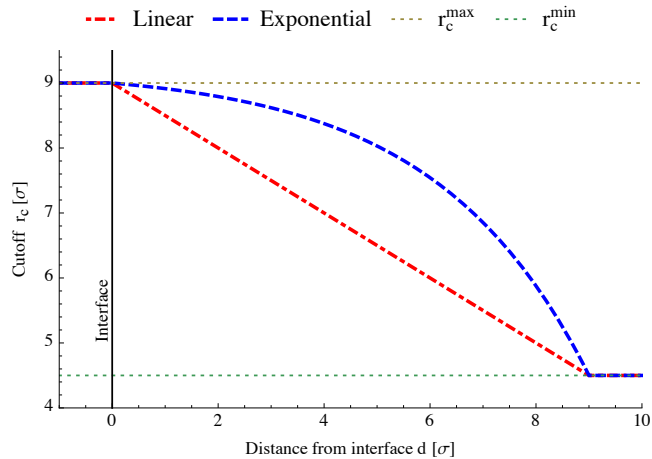

Figure 12: Exemplary cutoff functions for a minimum cutoff of $4.5\sigma$ and a maximum cutoff of $9.0\sigma$.

The cutoff function calculates the cutoff of each particle with respect to the particle-interface distance $d_i$. Two examples of such functions are given in Figure 12. Measurements of the *root-mean-square* (RMS) error of the forces (see Section 6.1) suggest that there exists an optimal function such that the error close to the interface is of the same order as the error within the interior. The choice of the cutoff function influences both the runtime as well as the accuracy of the method. The effect of the cutoff function on the RMS error of the forces can be found in Section 6.1.1.

The "exponential" cutoff function, shown in Figure 12 and used in the remainder of this thesis, is

defined as

$$r_c(d) = \begin{cases} r_c^{max}, & \text{if } d \leq 0 \\ a * 2.0^{\alpha\,(r_c^{max}-d)} + b, & \text{if } 0 \leq d < r_c^{max} \\ r_c^{min}, & \text{if } d > r_c^{max} \end{cases}, \quad (7)$$

where

$$a = \frac{r_c^{max} - r_c^{min}}{2^{\alpha\,r_c^{max}} - 1.0},$$
$$b = r_c^{min} - a$$
(8)

The parameter $\alpha$ determines the steepness of the function; the more negative $\alpha$ becomes, the more step-like the function gets (e.g., $\alpha = \infty$ yields a step function); we use $\alpha = -0.5$. $d$ represents the distance to the interface; we distinguish between particles inside the high-density phase (e.g., liquid phase) $d > 0$ and particles inside the low-density phase $d < 0$ (e.g., vapor phase). All particles at the interface or within the low-density phase are assigned the maximum cutoff $r_c^{max}$, while all particles which are at least $r_c^{max}$ away from the interface will be assigned the minimum cutoff $r_c^{min}$. Moreover, particles within the liquid phase will be assigned a progressively larger cutoff as they approach the interface. Although Equation 7 only suggests three cases, we could also imagine a function which applies the idea of a dynamic cutoff to those particles within the low-density phase as well and therefore require more cases.

3.2. Neighbor-list Build and Force Calculation

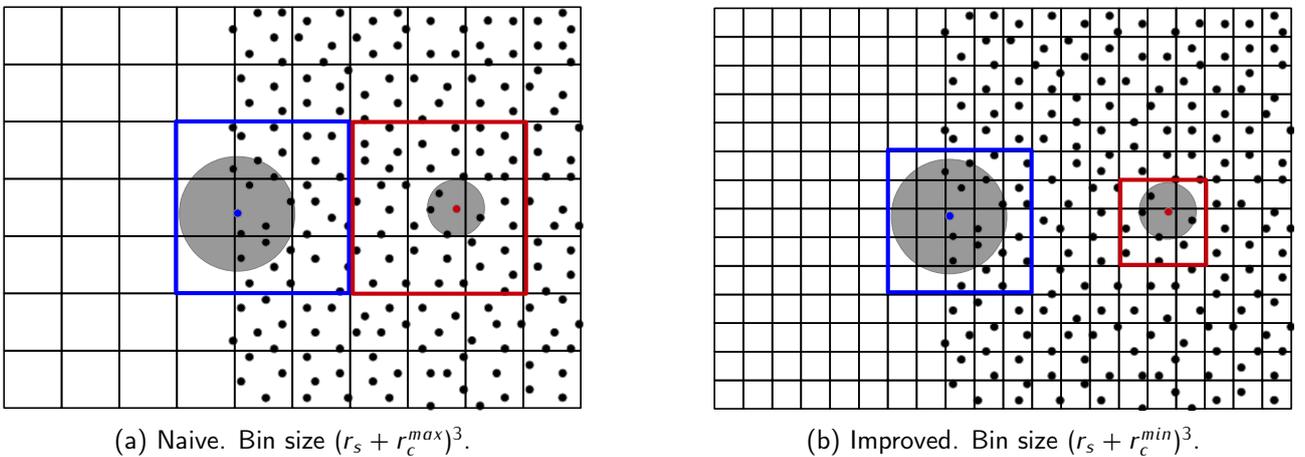

(a) Naive. Bin size $(r_s + r_c^{max})^3$.

(b) Improved. Bin size $(r_s + r_c^{min})^3$.

Figure 13: Neighbor-list stencil. The blue and red boundaries mark the volume that needs to be traversed for the blue or red particle, respectively.

The varying cutoffs pose additional challenges with respect to an efficient implementation of the neighbor-list build. A naive implementation which uses bins of size $r_s + r_c^{max}$ (see Figure 13a) would result in poor neighbor-list performance because particles with small cutoffs would still have to traverse the whole volume of $(3(r_s + r_c^{max}))^3$ to find all the interacting particles. This approach is especially bad for interior particles. For instance, if we use $r_c^{min} = \frac{1}{2}r_c^{max}$, particles within the interior would exhibit $\approx$ 98.5% spurious distance calculations which is far worse than the 84.5% which we have seen in Section 2.2.1. Given that the neighbor-list build is a memory-bound operation, this naive implementation would result in the same performance as if we did not use a dynamic cutoff at all.

The solution to this problem is the following: instead of using bins with edge length $r_s + r_c^{max}$, we use an edge length of $r_s + r_c^{min}$ or smaller (see Figure 13b). Thus, while we have to traverse a slightly larger stencil of neighboring boxes, the traversed volume is much smaller and results in fewer spurious distance calculations. Figure 13 illustrates this change. Notice that even the neighbor-list volume of interfacial particles decreases. This optimization yields a $4 - 6\times$ speedup of the neighbor-list build over the naive implementation.

The force calculation (see Listing 3.2) and 3.3) are very similar to the methods of Section 2.2.1 (compare Listing 2.1 and 2.2). The main difference is that the DCM versions use a dynamic cutoff per particle instead of a global, static cutoff; moreover, Newton's third law is no longer applicable. To be precise, while computing the force between particle $i$ and $j$ the force contribution to particle $j$ is no longer used; resulting in twice as many calculations. Note that the "magic" of the improved neighbor-list build is hidden in Line 4 of Listing 3.3.

**Algorithm 3.2** DCM. Force calculation
1: **for all** atoms $i$ **do**
2:     $f_i \leftarrow 0$
3:     **for all** neighbors $k$ of $i$ **do**
4:         $j \leftarrow \text{nbrs}_i[k]$
5:         $r_{ij} \leftarrow r_i - r_j$
6:         **if** $|r_{ij}| < r_c[i]$ **then**
7:             $f \leftarrow \text{forceLJ}(|r_{ij}|)$
8:             $f_i \leftarrow f_i - f \times r_{ij}$
9:         **end if**
10:     **end for**
11:     $\text{force}[i] \leftarrow \text{force}[i] + f_i$
12: **end for**

**Algorithm 3.3** Verlet-list build
1: **for all** local atoms $i$ **do**
2:     $\text{nNbrs}[i] \leftarrow 0$
3:     $\text{iBin} \leftarrow \text{getBin}(i)$
4:     **for all** $\text{jBin} \in \text{neighbors(iBin)}$ **do**
5:         **for all** atoms $j$ of jBin **do**
6:             $r_{ij} \leftarrow r_i - r_j$
7:             **if** $|r_{ij}| < r_c[i]$ AND $i \neq j$ **then**
8:                 $\text{nbrs}_i[\text{nNbrs}[i]] \leftarrow j$
9:                 $\text{nNbrs}[i] \leftarrow \text{nNbrs}[i] + 1$
10:             **end if**
11:         **end for**
12:     **end for**
13: **end for**

## 4. Interface Detection

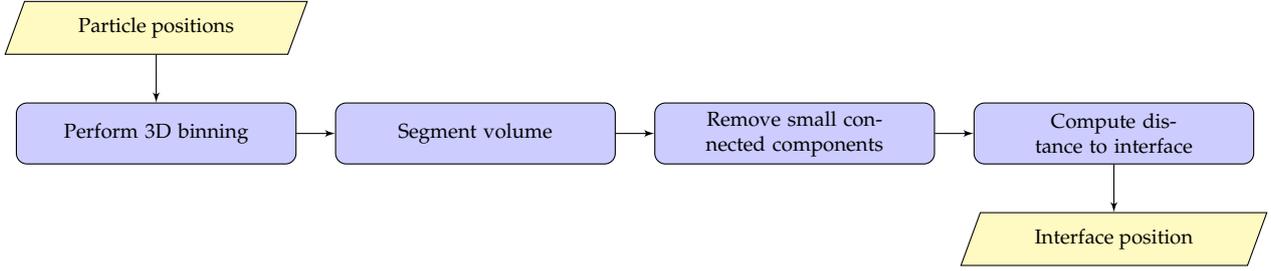

Figure 14: Interface detection.

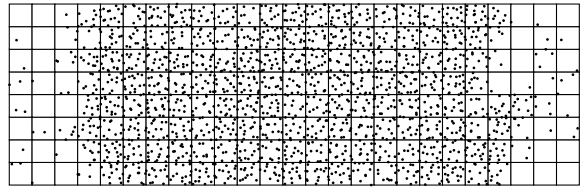

Figure 15: Binning.

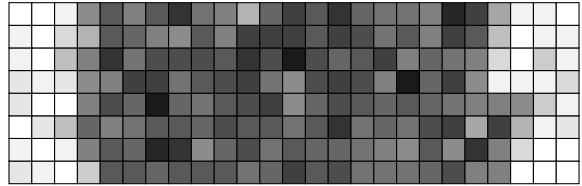

Figure 16: Create gray-scale image/volume.

Figure 14 shows a schematic of our interface detection method. The first step is to bin all particles into 3-dimensional boxes $b_{i,j,k}$ of size $b_x \times b_y \times b_z$ (see Figure 15) and treat the set $B$ of these "density values" as a 3-dimensional image of gray-scale values (see Figure 16),

$$B = \{b_{i,j,k} \in \mathbb{N} \mid 0 \leq i < N_x, 0 \leq j < N_y, 0 \leq k < N_z\}, \tag{9}$$

where $N_x$, $N_y$ and $N_z$ are integers computed according to Equation 10:

$$N_x = L_x/b_x, \quad N_y = L_y/b_y, \quad N_z = L_z/b_z, \tag{10}$$

such that $b_x \approx b_y \approx b_z$; and $0 < \alpha \leq b_x, b_y, b_z \leq \beta$. The resolution of the surface is determined by the parameter $\beta$; smaller values of $\beta$ result in a higher resolution. Since the runtime of this algorithm is determined by $N_x, N_y$, and $N_z$, $\beta$ influences the runtime as well. For obvious reasons there are limitations on the choice of both $\alpha$ and $\beta$. For instance, if $b_x, b_y$, or $b_z$ become too small (e.g., less than $\sigma$) it becomes highly unlikely that a box will have more than a single particle; moreover, many boxes will be empty and the segmentation algorithm (see Section 4.1) will fail. If, on the other hand, $b_x, b_y$ and $b_z$ become too large, the grid becomes too coarse and the interface cannot be precisely represented.

The optimal values for $b_x, b_y$, and $b_z$ depend on the average distance between two particles and thus on the system under investigation. A good choice for these values is given by the following rule of thumb: make $b_x$, $b_y$ and $b_z$ small enough to capture the interface to a desired accuracy and large enough that all boxes within the liquid phase have roughly the same number of particles (i.e., they should be large enough such that density fluctuations within the liquid phase are mitigated). For our LJ system we used a value of $b_x \approx b_y \approx b_z \approx 2.6\sigma$.

The current implementation is restricted to the detection of interfaces of systems with just two different densities (e.g., water in the liquid and vapor phase) but an extension to multicomponent systems is easily accommodated.

The proceeding subsections discuss the remaining tasks in Figure 14.

## 4.1. Segmentation

Once the set $B$ (see Equation 9) has been computed we have to segment this image/set into a segmented set $S$

$$S = \{s_{i,j,k} \in \{0,1\} \mid 0 \leq i < N_x, 0 \leq j < N_y, 0 \leq k < N_z\}, \quad (11)$$

$$\text{such that: } s_{i,j,k} = \begin{cases} 0, & \text{if } b_{i,j,k} \approx \rho_l, \\ 1, & \text{if } b_{i,j,k} \approx \rho_v \end{cases}, \quad (12)$$

where $\rho_l$ and $\rho_v$ are the average liquid and vapor densities, respectively. Boxes with a density closer to the liquid density should thus be labeled with 0, while boxes with a density closer to the vapor density should be labeled with 1. (see Figure 17).

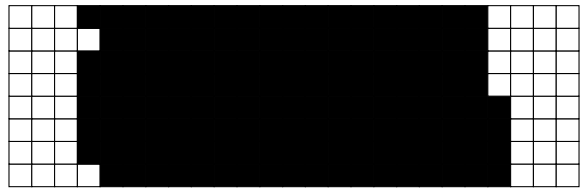

Figure 17: Segmented 2D imgage. Black boxes denote liquid phase; white boxes denote vapor phase.

The segmentation can be computed as the minimization of the piecewise constant Mumford-Shah functional for two-phase segmentation [28, 40]:

$$E[V^L, \rho_v, \rho_l] = \int_{V^L} (\rho(x) - \rho_v)^2 dx + \int_{V^V} (\rho(x) - \rho_l)^2 ds + \nu Per(V^L) \quad (13)$$

where $V^L, V^V \in V$ denote the liquid and vapor volumes, respectively, and $Per(V^L)$ is the perimeter of set $V^L$. The minimization of this functional yields the segmented volumes $V^L$ and $V^V$.

The parameter $\nu \in \mathbb{R}^+$ can be seen as a penalty for a large perimeter. It determines the smoothness of the surface (see Figure 18). If $\nu$ is too small, the interfaces will not be properly detected (see Figure 18a). If, on the other hand, $\nu$ is too large the penalty will become too large and all boxes will belong to the same set; hence, no interface will be detected (not shown). For our simulation and accuracy demands, we empirically determined $\nu = 0.005$. Even though we did not experience any problems with this value for any of our systems, automatic selection of $\nu$ remains an open question.

The implementation of this segmentation method is available in the open-source library QuocMesh

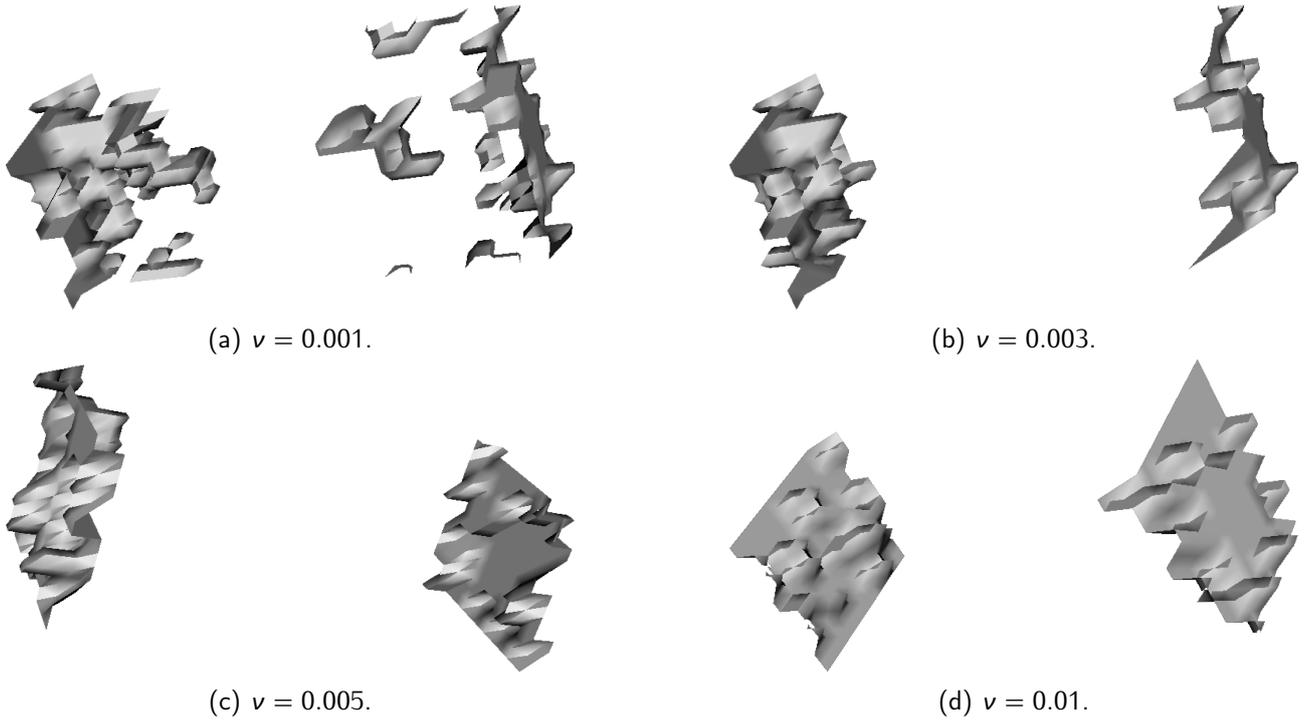

(a) $\nu = 0.001$.

(b) $\nu = 0.003$.

(c) $\nu = 0.005$.

(d) $\nu = 0.01$.

Figure 18: Effect of $\nu$ on the smoothness of the interface. The interfaces shown represent the liquid/vapor interface of a LJ system with 16000 particles after one million timesteps at a rather high temperature.

[29]. However, the QuocMesh implementation was not suited for periodic boundary conditions (PBC), so we had to extend QuocMesh's functionality. Most of the runtime of segmentation is spent on calculating finite differences. This makes a MPI parallelization straightforward, Using domain decomposition and the concept of ghost-cells, and will not be discussed in this thesis.

**Automatic detection of density values.** To make the interface detection as automatic as possible we need to find good estimates for $\rho_v$ and $\rho_l$. Naive estimates for $\rho_l$ and $\rho_l$ would be $\rho_v = min_{i,j,k} b_{i,j,k}$ and $\rho_l = max_{i,j,k} b_{i,j,k}$. However, this does not always give good results since the minimum and maximum might have large variances. Moreover, poorly chosen $\rho_l$ and $\rho_v$ values can result in totally useless results. Hence, it is desirable to detect these values automatically and with high precision.

One solution to this problem is the *K-Means* clustering algorithm [41, 42]. *K*-Means tries to minimize the following objective function:

$$E = \sum_{j=1}^{K} \sum_{i=1}^{\widetilde{N}} ||b_i - c_j||^2, \qquad (14)$$

where $c_j$ denotes a cluster center, $K$ the total number of clusters and $\widetilde{N} = N_x \times N_y \times N_z$ the total number of boxes. In our case $K = 2$, $c_0 \equiv \rho_l$ and $c_1 \equiv \rho_v$. Figure 19 shows the histogram over all boxes $b_{i,j,k}$ of a system with 16000 particles after one million timesteps at a high temperature. Figure 19 can be interpreted as follows: bars to the left of the plot denote low density areas (i.e., vapor phase); bars to the right of the plot denote high density areas (i.e., liquid phase) and boxes in between represent interface densities. The vertical, red lines show that *K*-means automatically detects estimates for $\rho_l$ and $\rho_v$.

## 4 INTERFACE DETECTION

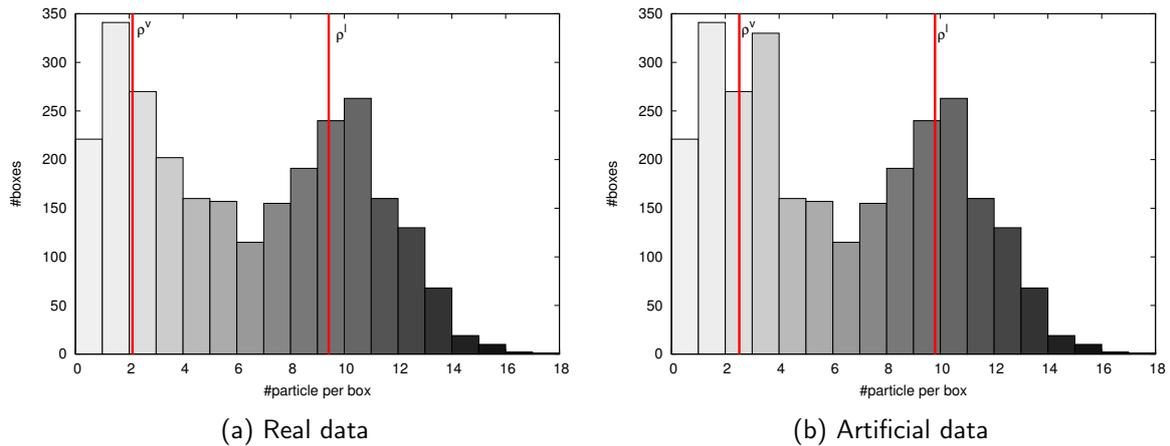

(a) Real data      (b) Artificial data

Figure 19: Histogram and k-means results. Red lines denote the clusters found by k-means.

### 4.2. Removal of Small Connected Components

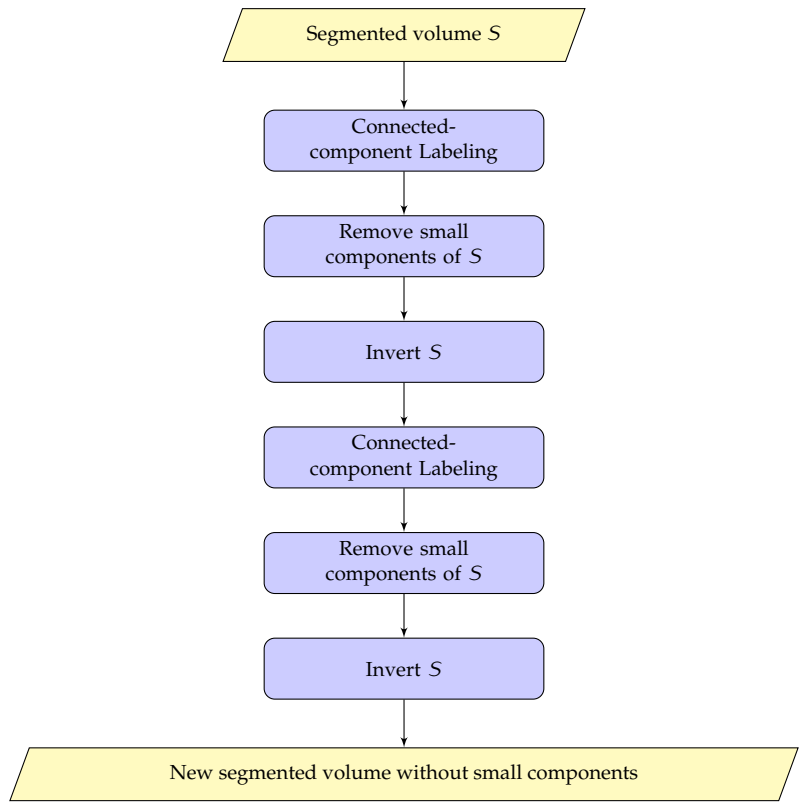

Figure 20: Outline for the removal of small connected-components.

Depending on the choice of $v$ and the accuracy of the segmentation algorithm small *bubbles* within the vapor phase may be falsely identified as liquids (or *vice versa*). Figure 21a shows an artificial example of such a case. To remove all small components (or bubbles) smaller than a certain volume, we apply a connected-component labeling algorithm [43] to find these connected components and their volumes. The outline of this algorithm is depicted in Figure 20. The two boxes labeled with *remove small components of S* remove all bubbles within the liquid phase and vapor phase, respectively. This technique could also be used to determine the total number of connected components or restrict this number to a user-specified value.

Our 3D implementation of the connected-component labeling algorithm is based on the existing 2D implementation already available within the QuocMesh library [29]. We extended this algorithm to 3D, incorporated PBCs, and parallelized it for distributed-memory systems (see Section 5.2). The existing implementation (see Algorithm 4.1) within QuocMesh is itself based on [44].

---

**Algorithm 4.1** Connected-component Labeling

```
1:  for 0 ≤ j < N_y do                              ▷ Loop over the image in row-major order
2:      for 0 ≤ i < N_x do
3:          if s_{i,j} == 0 then                    ▷ Only consider pixels of one phase
4:              nlNeighbors ← getNumLabeldNeighbors(i, j)
5:              if nlNeighbors == 0 then
6:                  label_{i,j} ← getNewLabel()
7:              else if nlNeighbors == 1 then
8:                  label_{i,j} ← getLabelOfNeighbor()
9:              else
10:                 label_{i,j} ← getLabelOfNeighbor()
11:                 markNeighboringLabelsAsEquivalent()
12:             end if
13:         end if
14:     end for
15: end for
16: mergeAllEquivalentLabels()
```

---

For better readability we restrict the description of the labeling algorithm to the two dimensional case. However, a 3D-extension is straightforward. A high-level overview of the connected-component labeling procedure is shown in Algorithm 4.1. The algorithm loops over all pixels $s_{i,j}$ (the output of the segmentation algorithm) in row-major order and assigns each pixel a label according to the following rules:

- If a pixel has no neighbor with a label, a new label will be assigned.

- If a pixel has exactly one neighbor, the label of the neighbor will be assigned.

- In case multiple neighbors have a label, the smallest label will be assigned and all neighboring labels will be marked as equivalent.

The details hidden in Line 11 and 16 will not be discussed further because these are technical details not required to understand for the parallel implementation discussed in Section 5.2. Figure 21 illustrates the steps necessary to remove all small bubbles for an artificial segmented image.

### 4.3. Compute Distance to Interface

Once we have removed all small components in $S$, we need to compute the distances $d_{i,j,k}$ to the interface of the vapor phase set $S^v \subseteq S$ and the liquid phase set $S^l \subseteq S$ with $S^v \cap S^l = \emptyset$. Since our current implementation only applies the dynamic cutoff method to particles within the liquid phase (compare Equation 7) we can set $d_{i,j,k} = 0$ (see Equation 6) when $s_{i,j,k} = 1$ holds. The remaining distances can be computed by solving the signed distance function which satisfies the Eikonal equation [24]:

$$|\nabla d(x)| = 1, x \in \mathbb{R}^n,$$
$$d(x) = 0, x \in \Gamma \subset \mathbb{R}^n \quad (15)$$

where $d(x)$ and $\Gamma$ represent the distance function and the interface, respectively.

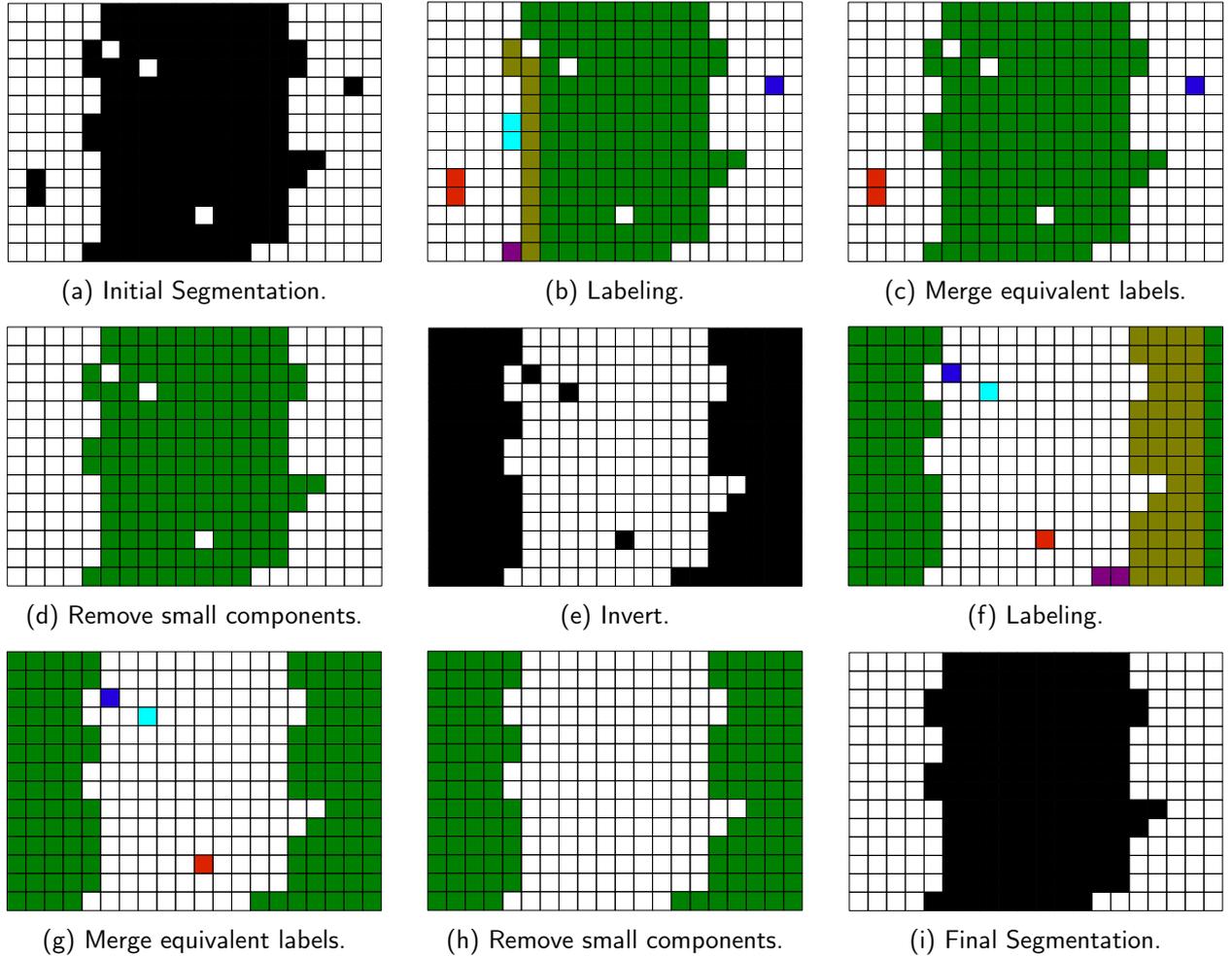

Figure 21: Removal of small connected-components under periodic boundary conditions.

The *fast marching method* (FMM) [45] solves this equation in $\mathcal{O}(\widetilde{N} \log \widetilde{N})$ time with respect to the total amount of grid points $\widetilde{N}$. Zhao et al. introduced an alternative solution to the Eikonal equation called the *fast sweeping method* (FSM) [24] as well as provided a detailed analysis of this algorithm, including error bounds [46]. The FSM has a linear-time complexity $\mathcal{O}(\widetilde{N})$ and is our preferred choice for the dynamic cutoff method. The remainder of this section outlines the original FSM as described in [46] and then introduces our own version which makes some improvements.

4.3.1. Fast Sweeping Method

The fast sweeping method is mainly based on upwind schemes and a Gauss-Seidel solver using different sweeping directions [24]. For *n* dimensions, Zhao et al. showed that $2^n$ Gauss-Seidel iterations with alternating sweeping orderings are enough to give a good approximation to the distance function for arbitrary data sets [46].

Algorithm 4.2 outlines the implementation of FSM for two dimensions (a 3D implementation with PBC is straightforward). *D* represents the grid of distances, and up-right, up-left, down-right and down-left denote the sweeping directions.

The discretization of Equation 15 using a Godunov upwind scheme results in the following non-linear equation [47]:

$$[(d_{i,j} - d_{xmin})^+]^2 + [(d_{i,j} - d_{ymin})^+]^2 = h^2, \tag{16}$$
$$0 \leq i < N_x, \, 0 \leq j < N_y,$$

where *h* is the grid spacing,

$$d_{xmin} = min(d_{i-1,j}, d_{i+1,j}), \, d_{ymin} = min(d_{i,j-1}, d_{i,j+1}), \tag{17}$$

and

$$(x)^+ = \begin{cases} x, & x > 0 \\ 0, & x \leq 0 \end{cases} \tag{18}$$

The Gauss-Seidel iterations sweep over the entire grid *D* in alternating directions and solve the Equation 16 for each grid point. The solution of Equation 16 $\overline{d}_{i,j}$ is then used to update each grid point according to

$$d_{i,j}^{new} = min(d_{i,j}^{old}, \overline{d}_{i,j}). \tag{19}$$

Figure 22 illustrates the correctness of the fast sweeping method using a circle as an example. For all boxes to have the proper distances, we need a total of four sweeps. An up-right GS sweep yields the correct distances for those grid points marked with *up-right* (see Figure 22). After three additional sweeps—up-left, down-right, and down-left— all distances will be computed correctly.

**Algorithm 4.2** 2D Fast Sweeping Method
1: **for all** $0 \leq j < N_y$ **do**
2:     **for all** $0 \leq i < N_x$ **do**
3:         **if** $s_{i,j} == 1$ **then**
4:             $d_{i,j} \leftarrow 0$     ▷ Vapor phase
5:         **else**
6:             $d_{i,j} \leftarrow$ in     ▷ Liquid phase
7:         **end if**
8:     **end for**
9: **end for**
10: GaussSeidel( $D, N_x, N_y$, up-right)
11: GaussSeidel( $D, N_x, N_y$, up-left)
12: GaussSeidel( $D, N_x, N_y$, down-right)
13: GaussSeidel( $D, N_x, N_y$, down-left)

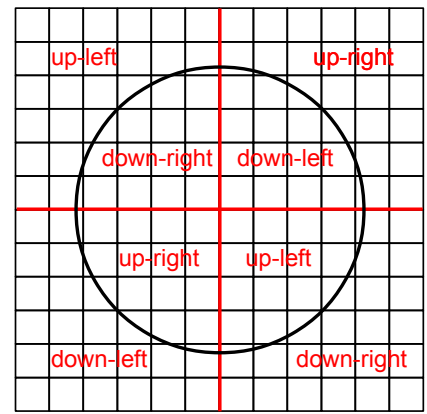

Figure 22: 2D FSM for a circle [46].

4.3.2. Cutoff-based Fast Sweeping Method

Problems arising from a parallel implementation of FSM (see Section 5.3) and the special needs of our distance calculations led us to develop a *cutoff-based fast sweeping method* (CFSM). The fundamental ideas behind CFSM are the following:

- Restrict distance calculations to those boxes close to the interface (i.e., $d_{i,j,k} < r_c^{max} + h$).

- Start computation of the distances at the interface.

The former idea is specially tailored for the DCM, which does not require particle-interface distances beyond a certain distance. The latter idea is more general and is also applicable to the original FSM. Algorithm 4.3 describes the implementation of CFSM.

---

**Algorithm 4.3** Cutoff-based fast sweeping method.

1: initialize($D^p$, $Q$)  ▷ Add interfacial boxes to $Q$
2: **for** $0 \leq \text{iter} < \text{iter}^{\max}$ **do**
3:     **for all** boxes $(x, y, z) \in Q$ **do**
4:         $d^{new} \leftarrow$ solveEikonal($D^p$, $(x, y, z)$)  ▷ local FSM
5:         **if** $d^{new} \leq r_c^{\text{grid}}$ **then**
6:             **if** $|(D^p)_{x,y,z} - d^{new}| > \Delta e$ **then**
7:                 $\Delta e \leftarrow |(D^p)_{x,y,z} - d^{new}|$
8:             **end if**
9:             $(D^p)_{x,y,z} \leftarrow d^{new}$
10:            addNeighborsToQueue($\widetilde{Q}$, $(x, y, z)$)
11:         **end if**
12:     **end for**
13:     swapQueues($Q$, $\widetilde{Q}$)
14:     emptyQueue($\widetilde{Q}$)
15:     **if** $\Delta e < \epsilon$ **then**
16:         **break**
17:     **end if**
18: **end for**

---

Figure 23 illustrates the propagation of the interface. Red dots mark those grid-points just updated, while non-updated grid-points remain black.

**Performance.**  Our numerical experiments show that CFSM and FSM give the same results for grid points within the cutoff (data not shown). Moreover, as is evident in Figure 24 CFSM is preferable over FSM for various inputs. The first six inputs of Figure 24 denote artificial examples with randomly placed points (i.e., $d_{i,j,k} = 0$) while the remaining two inputs are much more similar to a typical input from an MD simulation (i.e., consisting of contiguous regions). The performance of CFSM increased by roughly an order of magnitude over the original fast sweeping method for relevant MD inputs.

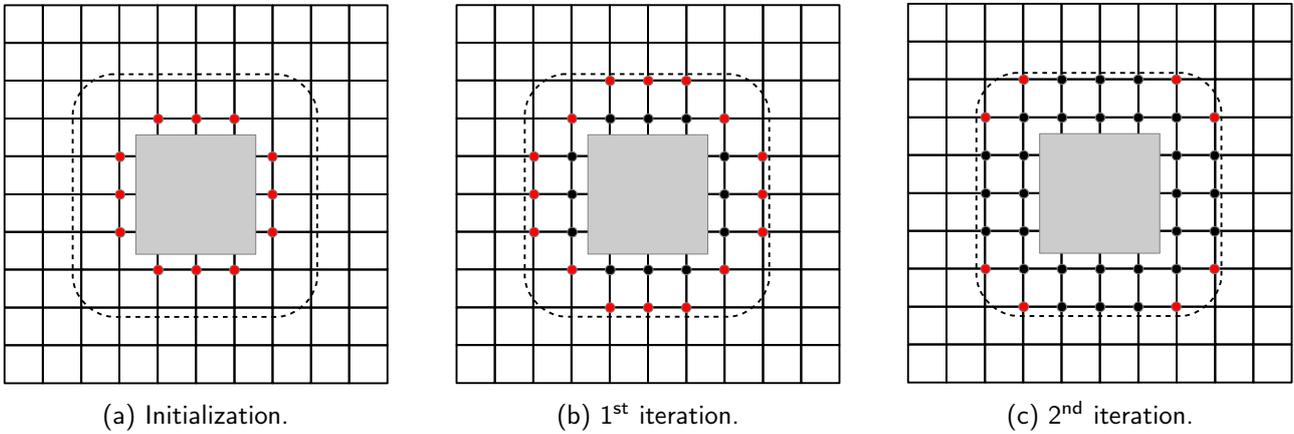

(a) Initialization.      (b) 1$^{\text{st}}$ iteration.      (c) 2$^{\text{nd}}$ iteration.

Figure 23: Cutoff-based FSM illustrated. Dashed line denotes the cutoff.

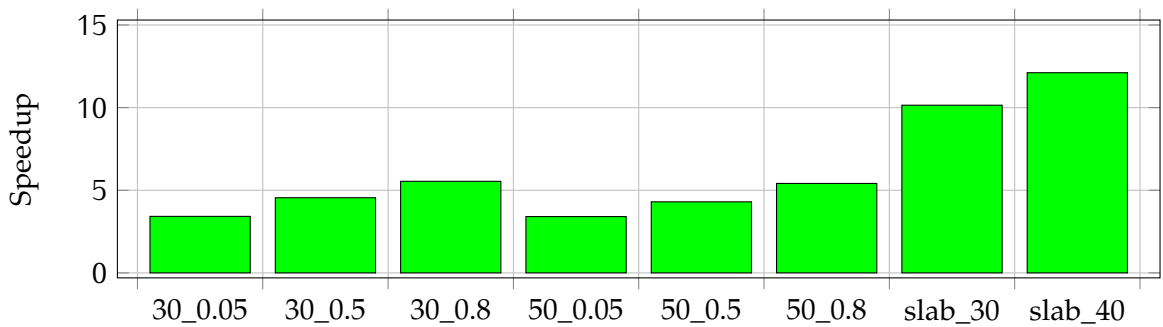

Figure 24: Speedup of CFSM over FSM for various examples using a cutoff of $4h$. Examples are encoded as follows: $x\_y$, x denotes a grid of size $x^3$ and y denotes the relative amount of randomly placed "interfacial" grid-points. The last two examples *slab_x* represent a domain of size $x^3$ with a continuous slab of width $\frac{x}{2}$.

## 5. Parallelization

Our parallel DCM implementation is based on spatial decomposition as this is the main parallelization technique used in LAMMPS [38]. This means that each process is responsible for a certain subdomain and all its local particles. Hence, each process only has a memory requirement of $\mathcal{O}(N/P)$, where $N$ and $P$ are the number of particles and processes, respectively.

Figure 25 outlines the distributed-memory implementation of the dynamic cutoff method; red boxes mark operations which require communication. DCM is designed to be highly scalable and to avoid non-local communication wherever possible.

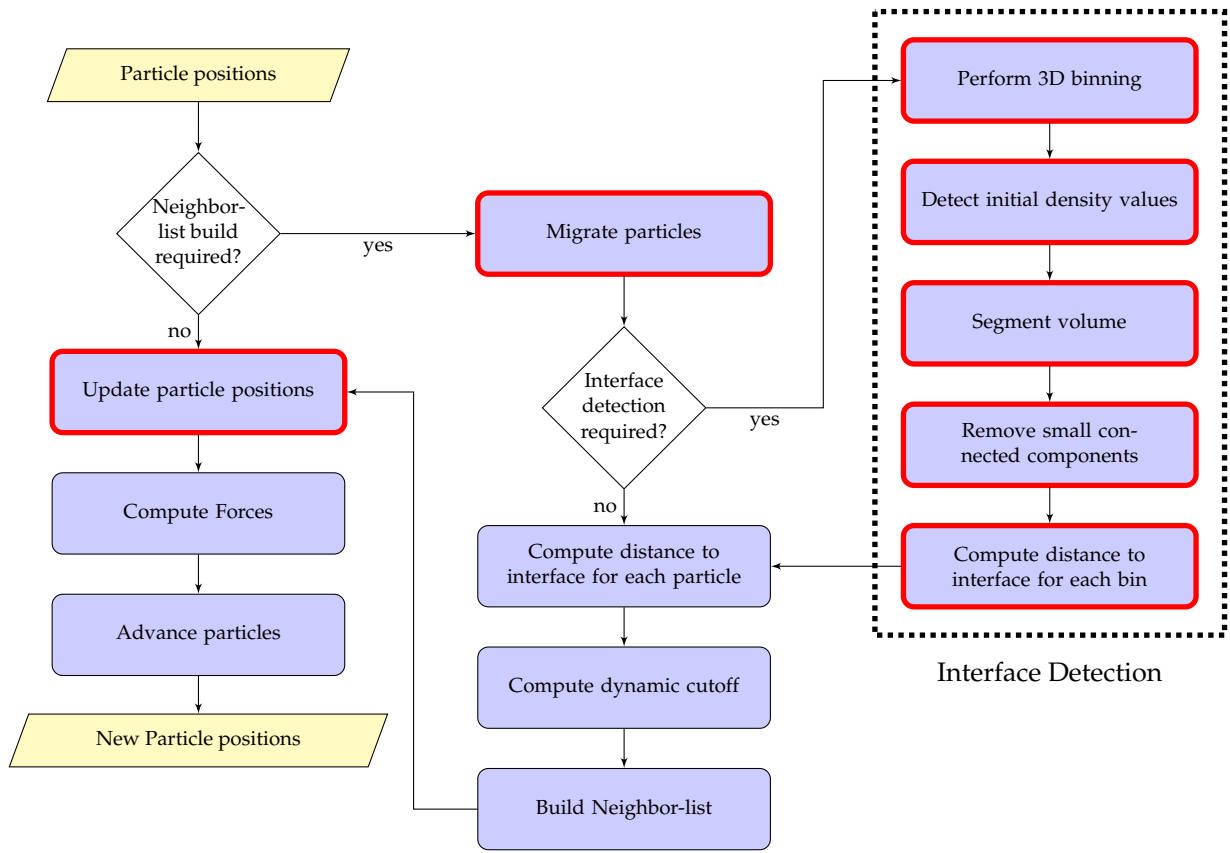

Figure 25: Breakdown of single DCM step.

The parallelization of the binning procedure, $K$-means and the segmentation algorithm will not be discussed in this thesis. Instead, we focus on the parallel 3D connected-component algorithm and the parallel CFSM.

We have also added a shared-memory parallelization based on OpenMP [48] in addition to our MPI implementation of DCM. This allows us to start multiple threads per MPI rank which reduces the memory requirements as well as the communication overhead and results in better performance (see Section 6.2.1).

### 5.1. Neighbor-list Build and Force calculation

The OpenMP implementation of the neighbor-list build and force evaluation is based on the USER-OMP package of LAMMPS. The shared-memory parallelization of the kernels for the force evaluation and the Verlet-list build (see Algorithms 5.1 and 5.2) only requires a single OpenMP directive in front of the outer-most loops[8].

---

**Algorithm 5.1** OpenMP DCM force calculation.
1: **#pragma omp for schedule**(dynamic,20)
2: **for all** atoms $i$ **do**
3:     $f_i \leftarrow 0$
4:     **for all** neighbors $k$ of $i$ **do**
5:       $j \leftarrow \text{nbrs}_i[k]$
6:       $r_{ij} \leftarrow r_i - r_j$
7:       **if** $|r_{ij}| < r_c[i]$ **then**
8:          $f \leftarrow \text{forceLJ}(|r_{ij}|)$
9:          $f_i \leftarrow f_i - f \times r_{ij}$
10:       **end if**
11:     **end for**
12:     $\text{force}[i] \leftarrow \text{force}[i] + f_i$
13: **end for**

**Algorithm 5.2** OpenMP DCM Verlet-list build.
1: **#pragma omp for schedule**(dynamic,20)
2: **for all** local atoms $i$ **do**
3:     $\text{nNbrs}[i] \leftarrow 0$
4:     $\text{iBin} \leftarrow \text{getBin}(i)$
5:     **for all** jBin $\in$ neighbors(iBin) **do**
6:       **for all** atoms $j$ of jBin **do**
7:          $r_{ij} \leftarrow r_i - r_j$
8:          **if** $|r_{ij}| < r_c[i]$ AND $i \neq j$ **then**
9:             $\text{nbrs}_i[\text{nNbrs}[i]] \leftarrow j$
10:             $\text{nNbrs}[i] \leftarrow \text{nNbrs}[i] + 1$
11:          **end if**
12:       **end for**
13:     **end for**
14: **end for**

---

We note here a small but important detail. Instead of using the default, static schedule, DCM requires a dynamic schedule with a chunk-size of about 20 particles. A static schedule (i.e., evenly distributing the particles among the threads) would result in severe load imbalancing because some particles have larger cutoffs than others and therefore require much more computational time. The dynamic schedule removes this problem by always assigning chunks of 20 particles at a time to each thread. Figure 26 shows the speedups of a dynamic schedule over the static schedule for both kernels separately. We observe that the force evaluation is more strongly affected by this change than the neighbor-list build.

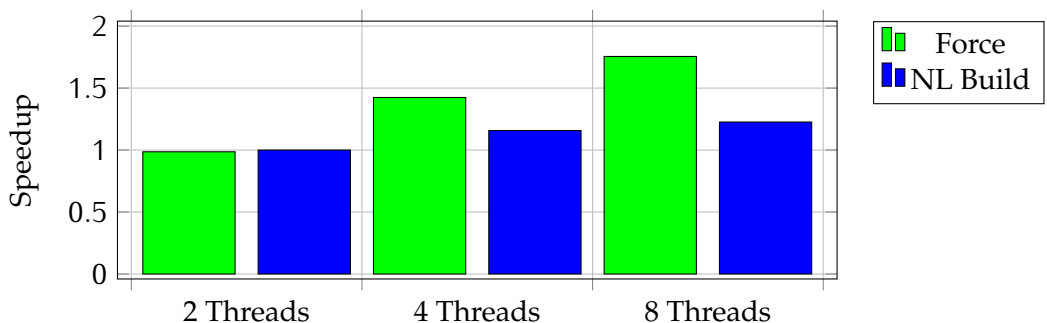

Figure 26: Speedup of dynamic schedule over static schedule running 1 MPI process with a 2, 4 or 8 threads. DCM with $r_c^{min} = 4\sigma$ and $r_c^{max} = 9\sigma$. Host: Intel Xeon CPU E5-2670.

---

[8]The actual implementation requires a few more tweaks but the concept remains the same.

## 5.2. Removal of Small Connected-Components

Our goal for a distributed-memory parallelization for the removal of small connected components (see Section 4.2) was to design a scalable algorithm with a strictly local communication pattern. This restriction, as we will see in this section, makes it necessary to content ourselves with an approximate solution sufficient for our purposes. To be precise, this method gives a lower bound on the actual *label sizes* (the size of each connected component) with only very little local communication.

Algorithm 5.3 gives a high-level description of the parallel labeling algorithm in two dimensions (the 3D implementation is similar). Each process $p$ starts to label its own subset $S_p \subset S$ using the non-parallelized labeling algorithm from section 4.2 without applying PBCs. The next step is to communicate the boundary labels to the left, right, top and bottom neighbors, including PBCs. During this step equivalence classes are built and the labels are updated by summing up their sizes. Due to the fact that we prescribe an order on the communication—first left to right, then top to bottom— information flows first horizontally, then vertically. Hence, we need an additional local communication in the reverse direction (top-to-bottom) while using the maximum of the previously computed label sizes. Once this is done each process can independently remove those labels smaller than a certain threshold.

---

**Algorithm 5.3** Parallel removal of small components in 2D.

1: $L \leftarrow \text{label}(S_p)$ ▷ Serial labeling
2:
3: sendBoundary($L$, left-right) ▷ Non-blocking
4: $B \leftarrow$ recvBoundary() ▷ Blocking
5: $L \leftarrow$ updateBoundary($L$, $B$, SUM)
6: MPI_Barrier() ▷ Global synchronization
7:
8: sendBoundary($L$, top-bottom) ▷ Non-blocking
9: $B \leftarrow$ recvBoundary() ▷ Blocking
10: $L \leftarrow$ updateBoundary($L$, $B$, SUM)
11: MPI_Barrier() ▷ Global synchronization
12:
13: sendBoundary($L$, left-right) ▷ Non-blocking
14: $B \leftarrow$ recvBoundary() ▷ Blocking
15: $L \leftarrow$ updateBoundary($L$, $B$, MAX)
16:
17: $S_p \leftarrow$ removeSmallComponents($S_p$, $L$)

---

Figure 27 illustrates the steps of the algorithm on an artificial example. After the first communication (see Figure 27c) neighboring processes will have detected equivalent labels and updated their sizes according to their left and right neighbors. Figure 27d shows the labels and label sizes after the communication with the top and bottom neighbors. Please note that this is not sufficient to compute the label sizes correctly (marked with red circles). Another communication step in the reverse direction eliminates this problem (see Figure 27e).

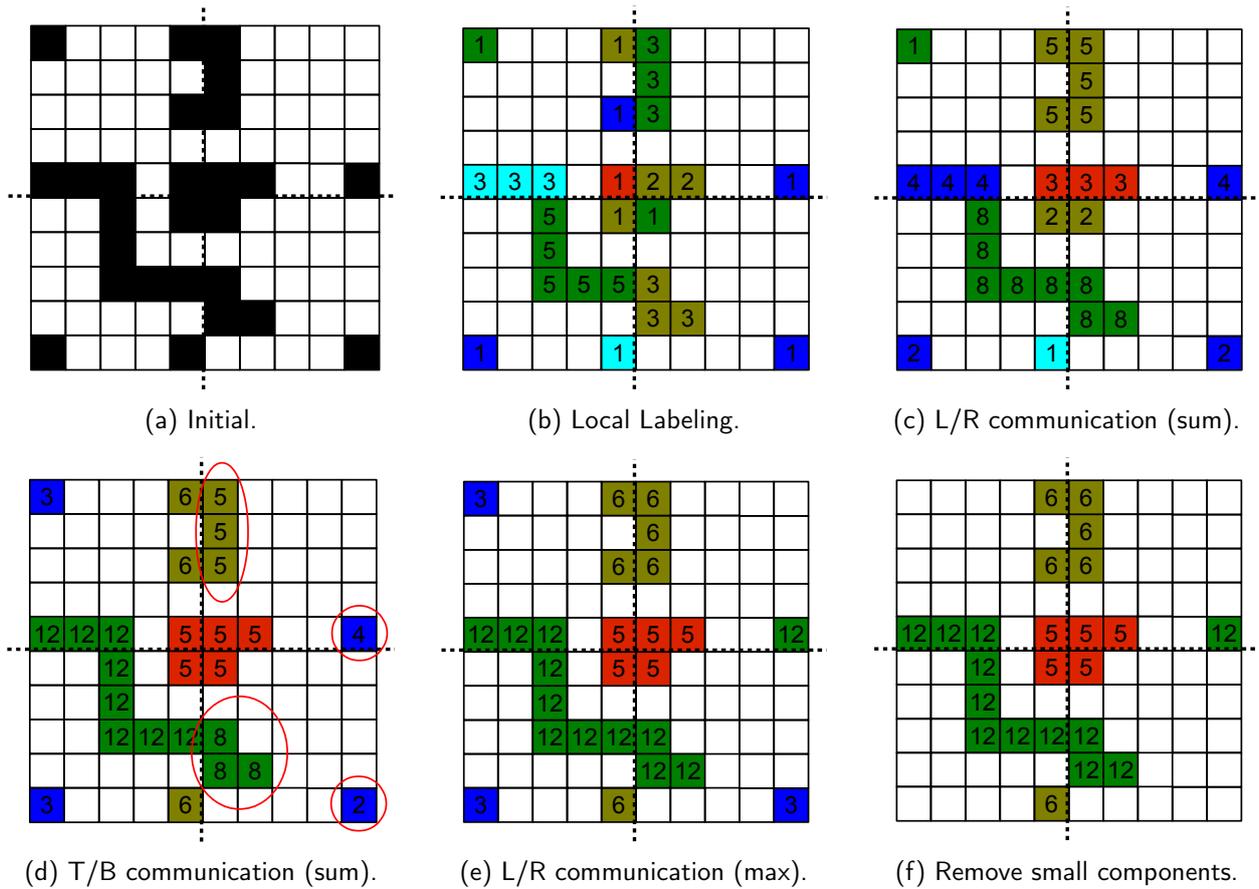

(a) Initial.　　(b) Local Labeling.　　(c) L/R communication (sum).

(d) T/B communication (sum).　　(e) L/R communication (max).　　(f) Remove small components.

Figure 27: Parallel removal of small components using PBCs. All components smaller than 5 are removed. Numbers denote the component size.

We note that there exist pathological cases for which this algorithm does not give the correct solution; Figure 28 illustrates such a case. However, these cases are highly unlikely to occur in MD simulations. Moreover, this algorithm would still yield good lower bounds of the component sizes. These lower bounds suffice for our purposes, since we are interested in removing **small**[9] components. As an example, a label spread among multiple processors (see Figure 28) is typically large enough that our communication pattern suffices to increase the label size above the selected threshold.

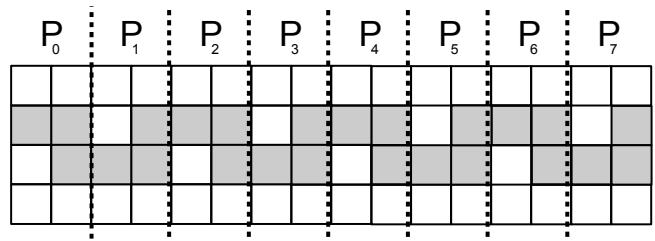

Figure 28: Pathological example which illustrates that local communication does not suffice to get the correct component size.

**Performance.** Figure 29 shows the speedup of the parallel labeling algorithm over its sequential counter part. Section 6.2 gives a detailed performance profile for the entire dynamic cutoff method and puts the performance of this labeling algorithm into perspective.

---

[9]If we were interested in large components this algorithm would not be nearly as good.

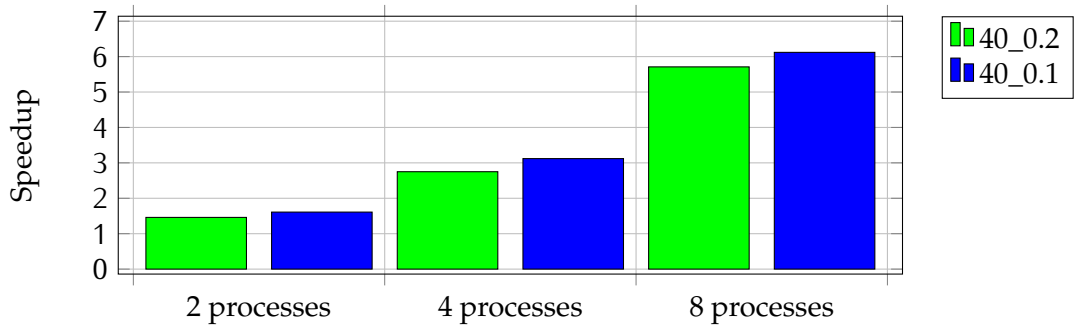

Figure 29: Speedup of parallel connected component labeling algorithm over serial implementation. Examples are encoded as follows: x_y, x denotes a grid of size $x^3$ and y denotes the relative amount of randomly placed "wholes" (e.g., $y = 0$ corresponds to a single large component without "wholes").

5.3. Cutoff-based Fast Sweeping Method

Zhao et al. [47] propose two different parallelization schemes for their fast sweeping method: (1) Parallelize over the different sweeping directions and (2) subdivide the domain evenly among the processes and apply multiple sweeps to each subdomain before exchanging boundary information with neighboring processes. While (1) does not scale beyond the number of sweeping directions, which is typically much less than the number of processes $P$, it also does not scale because every process requires $\mathcal{O}(\widetilde{N})$ memory. On the other hand, strategy (2) yields a much more scalable solution but might require many more floating-point operations than the serial version (see Appendix A). Moreover, their paper shows neither speedups nor scalability results.

The remainder of this section introduces the distributed-memory parallelization of our cutoff-based fast sweeping method. This parallel implementation is based on the principle of spatial decomposition. The underlying idea is simple: each process $i$ applies the serial CFSM to its subdomain $D_p$, exchanges the boundaries after every CFSM step and terminates once the maximum change $\Delta e$ is smaller than some threshold $\epsilon$. The outline for this parallelization is shown in Algorithm 5.4.

5 PARALLELIZATION

**Algorithm 5.4** Parallel CFSM.

1: initialize($D^p$, $Q$)  ▷ Add interfacial boxes to $Q$
2: **for** $0 \leq \text{iter} < \text{iter}^{\max}$ **do**
3:     **for all** boxes $(x, y, z) \in Q$ **do**
4:         $d^{new} \leftarrow \text{solveEikonal}(D^p, (x, y, z))$  ▷ local FSM
5:         **if** $d^{new} \leq r_c^{\text{grid}}$ **then**
6:             **if** $|(D^p)_{x,y,z} - d^{new}| > \Delta e$ **then**
7:                 $\Delta e \leftarrow |(D^p)_{x,y,z} - d^{new}|$
8:             **end if**
9:             $(D^p)_{x,y,z} \leftarrow d^{new}$
10:            addNeighborsToQueue($\widetilde{Q}$, $(x, y, z)$)
11:         **end if**
12:     **end for**
13:     swapQueues($Q$, $\widetilde{Q}$)
14:     emptyQueue($\widetilde{Q}$)
15:     $\Delta e \leftarrow \text{MPI\_Allreduce}(\Delta e, \text{MAXIMUM})$
16:     **if** $\Delta e < \epsilon$ **then**
17:         **break**
18:     **end if**
19:     ghostExchange($D^p$)  ▷ Local communication
20:     addModifiedBoundariesToQueue($D^p$, $Q$)
21: **end for**

Figure 30 illustrates the successive steps of Algorithm 5.4 on an example.

**Performance.** The speedup of the parallel CFSM over the serial implementation is shown in Figure 31. The example was run on a 16-core Intel Xeon CPU E5-2670.

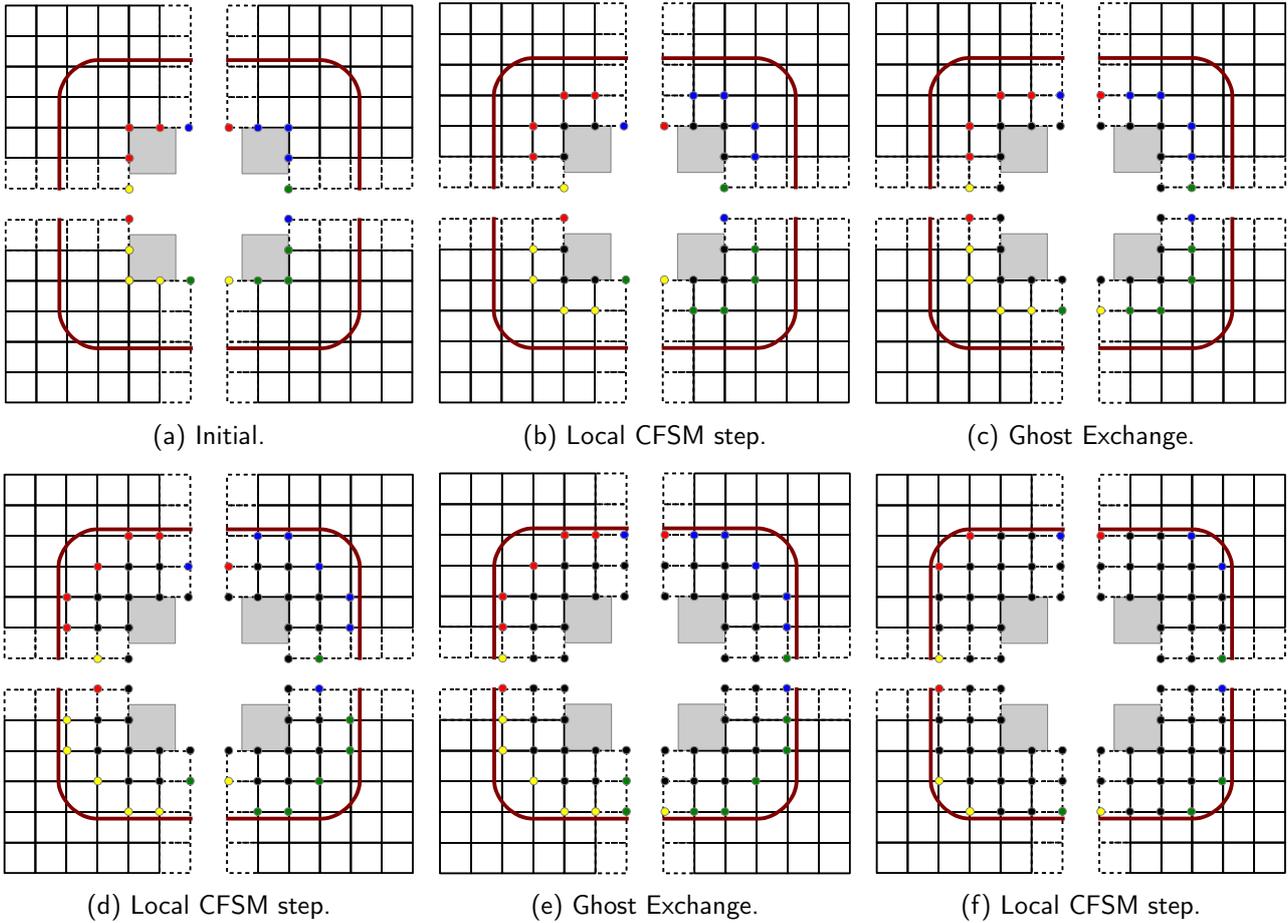

Figure 30: Parallel CFSM. Colored nodes denote active grid-points belonging to a certain processor. Black nodes denote inactive/complete grid-points. Dashed grid-cells denote ghost cells (i.e., cells belonging to a different process).

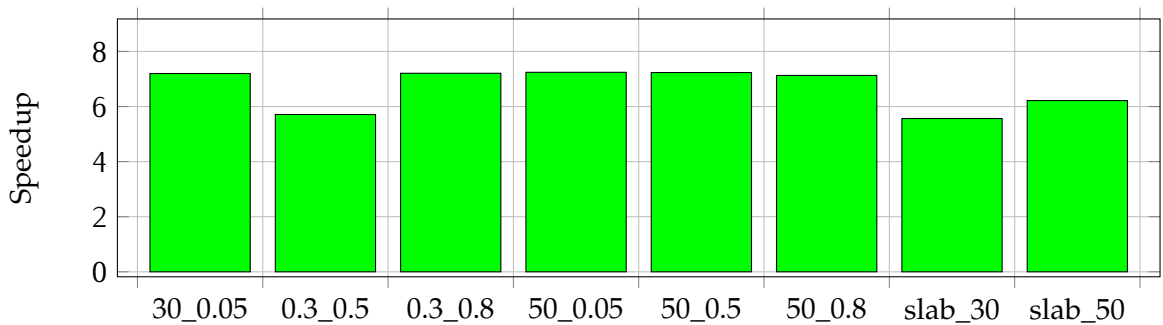

Figure 31: Speedup of parallel CFSM with 8 MPI processes over serial CFSM for various examples using a cutoff of $4h$. Examples are encoded as follows: $x\_y$, x denotes a grid of size $x^3$ and y denotes the relative amount of randomly placed "interfacial" grid-points. The last two examples *slab_x* represent a domain of size $x^3$ with a continuous slab of width $\frac{x}{2}$.

## 6. Results

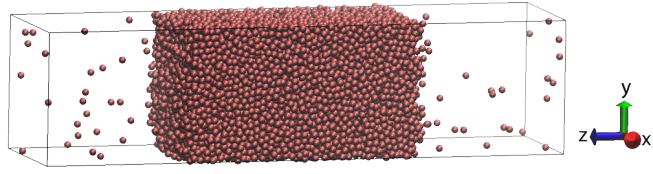

Figure 32: Interfacial system with 16000 Lennard-Jones particles in a domain of size $22\sigma \times 22\sigma \times 88\sigma$.

This section compares the accuracy and performance of DCM with both PPPM and the classical/static cutoff method. If not otherwise mentioned, we show results for a test system of 16000 LJ particles randomly placed in a $22\sigma \times 22\sigma \times 44\sigma$ box. This box is then centered in a $22\sigma \times 22\sigma \times 88\sigma$ domain (see Figure 32). The parameters are scaled to LJ units with temperature $T^* = k_B T/\epsilon = 0.7$ (i.e., well below the critical temperature), timestep $\tau = 0.005$ and distance $r^* = r/\sigma$. Moreover, all simulations are run with LAMMPS version *1-Feb-14* using an *NVT* ensemble (i.e., keeping the number of particles, the volume and the temperature constant). All measurements are taken after an equilibrating period of $100,000$ timesteps.

### 6.1. Accuracy

This section compares the accuracy of DCM against that of both PPPM and a static cutoff. Section 6.2 reuses these findings to perform a fair performance comparison.

Figure 33 illustrates that our method detects the interface at the proper location and adjusts the cutoff accordingly.

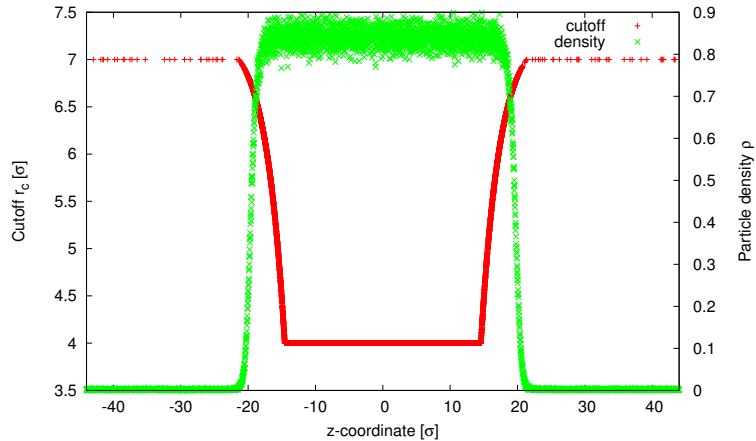

Figure 33: Dynamic cutoff ($r_c^{min} = 4.0\sigma$ and $r_c^{max} = 7.0\sigma$) superimposed on the particle density.

#### 6.1.1. Forces

We follow the same approach as Isele-Holder et al. [15] and measure the relative error in the forces according to:

$$\Delta F^{\text{rel}} = \sqrt{\frac{1}{N} \sum_{i=1}^{N} \left( \frac{F_i^{DCM} - F_i^{exact}}{F_i^{exact}} \right)^2}, \tag{20}$$

where $F_i^{exact}$ and $F_i^{DCM}$ respectively represent the exact force and the force calculated by DCM for particle *i* after a single timestep. We calculated $F_i^{exact}$ by using the Ewald summation with a large cutoff and a large number of reciprocal vectors to get a precise estimate of the exact solution.

Figure 34 shows the root-mean-square error in the forces for linear and exponential cutoff functions (see Section 3.1). We observe that:

1. the error in the forces decreases when increasing the minimal or maximal cutoffs;

2. increasing the minimum cutoff beyond $\approx 4.0\sigma$ hardly affects the accuracy;

3. the accuracy gain by increasing the maximum cutoff diminishes as the maximum cutoff increases; and

4. the exponential cutoff function yields a higher accuracy than the linear cutoff function.

Point (1) is somewhat obvious and shows that the algorithm works as expected: an increase in the computational demand corresponds to an increase in the accuracy. Points (2) and (3) suggest that there is an optimal combination of minimum and maximum cutoffs for a desired accuracy threshold $\Delta F^{\text{rel}}$; we investigate this later. To explain the contribution to the error for different parameter choices, a more elaborate analysis is required. Moreover, observation (4) is also expected because the "exponential" cutoff function assigns larger cutoffs to more particles than the "linear" cutoff function does and hence results in higher accuracy.

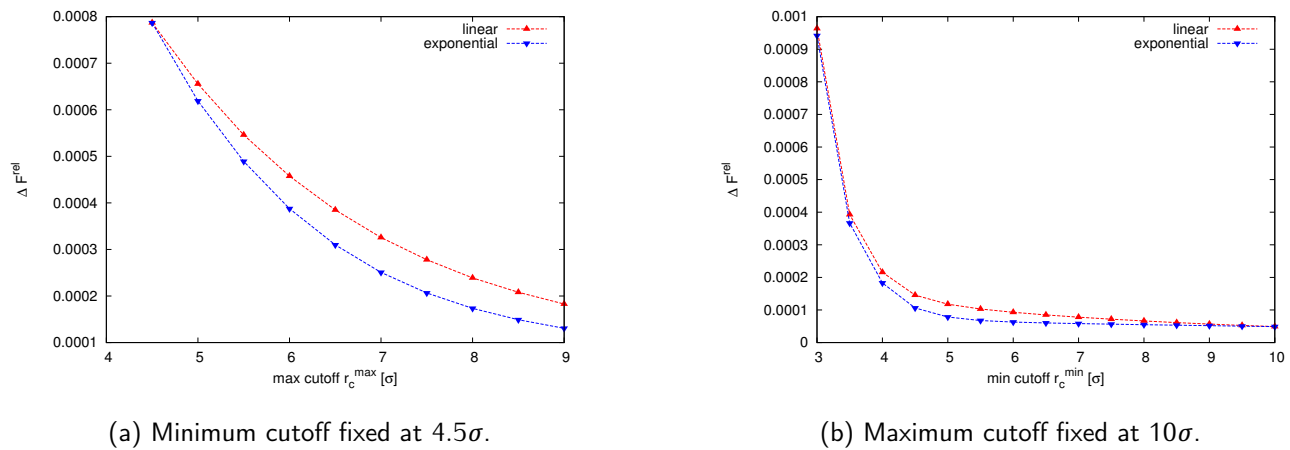

(a) Minimum cutoff fixed at $4.5\sigma$.  (b) Maximum cutoff fixed at $10\sigma$.

Figure 34: RMS error in the forces.

We measured the error of the forces $\Delta F_z$ in the *z*-direction perpendicular to the interface for different parameter choices (see Figure 35). Figure 35a shows $\Delta F_z$ for the classical cutoff method. We observe that the error decreases as the cutoff increases and that the error is clearly directed [49]; meaning that particles at the left interface, near $z \approx -21$, are missing a force "pulling" them to the right, while particles at the right interface, near $z \approx +21$, are missing a force "pulling" them to the left. It also highlights the main idea of the DCM since increasing the overall cutoff decreases the error at the interface significantly, while the effect on internal particles is moderate at best. Moreover, the error of the interfacial particles is roughly two order of magnitude larger than the error of the internal particles (i.e., $-10 \leq z \leq 10$); this demonstrates that a static cutoff is not the ideal choice for interfacial systems.

Figure 35b shows $\Delta F_z$ for DCM with varying maximum cutoffs and a linear cutoff function. It mainly indicates three things. First, as the maximum cutoff increases, the error for interfacial particles decreases. Second, interfacial particles are still the main source of error. Finally, a linear cutoff function is not optimal since the error increases with an increasing distance to the interface before it drops again. The latter observation led us to design of the "exponential" cutoff function in Equation 7 that reduces the error at the interface even further (see Figure 35c). A more step-like exponential function would reduce the error even further[10]. However, the design of an analytical cutoff function that respects the interaction potential is left for future work. The choice of the cutoff function not only influences the accuracy but also the performance so there might be an optimal choice for this function with respect to performance as well.

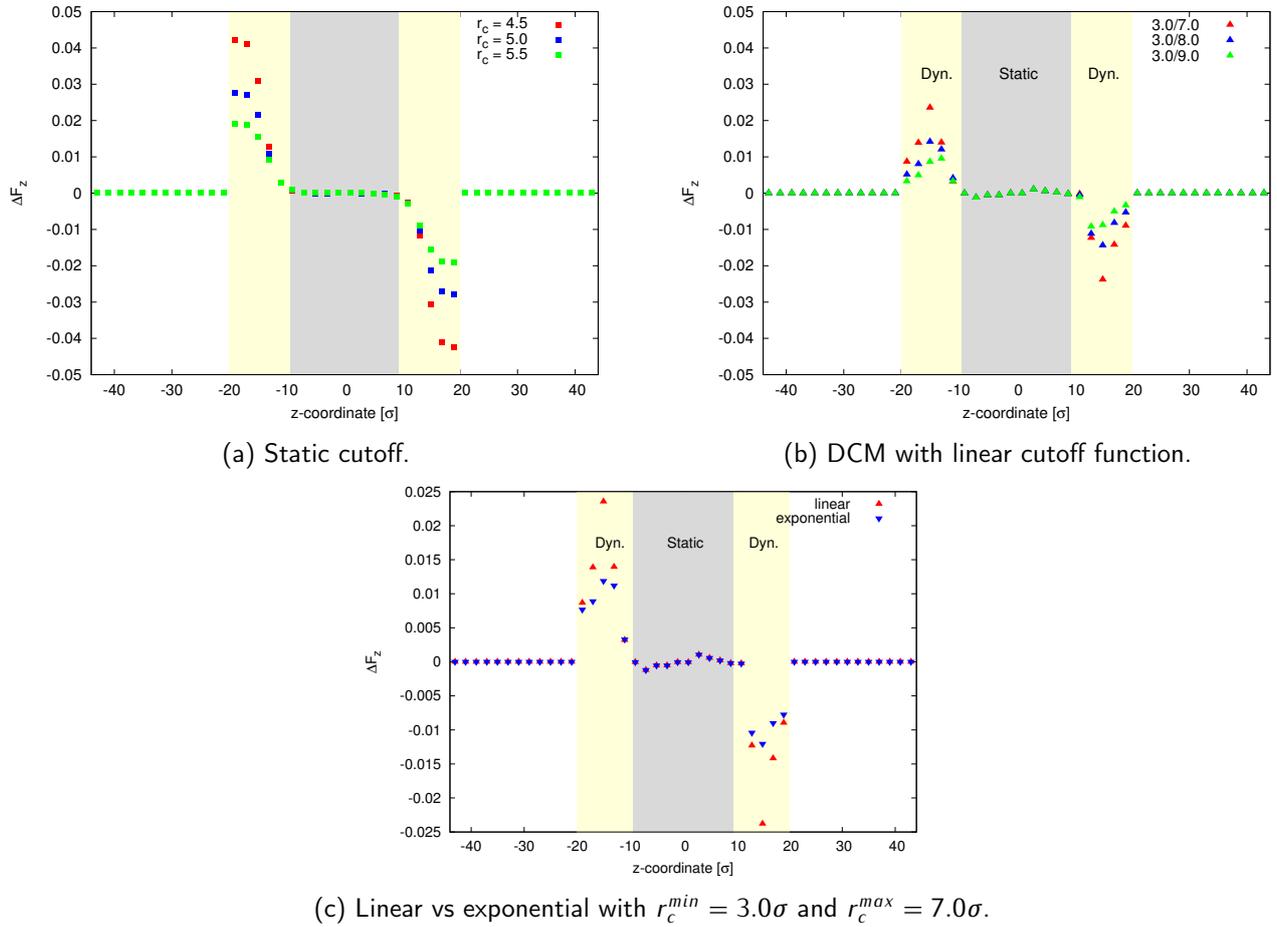

(a) Static cutoff.

(b) DCM with linear cutoff function.

(c) Linear vs exponential with $r_c^{min} = 3.0\sigma$ and $r_c^{max} = 7.0\sigma$.

Figure 35: Absolute error of the forces $\Delta F_z$. The yellow boxes denote the interfacial region while the gray box marks the internal region.

An important questions that still remains to be answered is how to choose the DCM parameters optimally. Figure 36 tries to answer this question. It shows the RMS error of the forces $\Delta F^{\text{rel}}$ with respect to both the minimal cutoff as well as the maximum cutoff. The origin of each arrow (see Figure 36) indicates the optimal parameter choice for a desired accuracy with respect to the best performance. For instance, to achieve an accuracy of at least $\Delta F^{\text{rel}} = 0.00149$, $r_c^{min} = 3.0\sigma$ and $r_c^{max} = 5.0\sigma$ would be optimal, while $r_c^{min} = 3.0\sigma$ and $r_c^{max} = 5.5\sigma$ would be the best choice for a slightly higher accu-racy of $\Delta F = 0.00133$. This figure validates our previous observation that the error monotonically decreases with an increasing minimum and maximum cutoff. The more important observation, however, is indicated by the arrows. The sequence in which to pick the optimal $r_c^{min}/r_c^{max}$ combination for a given accuracy is indicated by the arrow heads. We make the following observations:

1. The minimum cutoff has a stronger effect on performance[11] than the maximum cutoff. This is shown by the fact that the arrows point towards a larger maximum cutoff before pointing towards a larger minimum cutoff. In other words, with respect to performance it is preferable to increase the maximum cutoff instead of the minimum cutoff to achieve higher accuracy.

2. The optimal parameter choices are far away from the diagonal (the diagonal represents a static cutoff); hence, a dynamic cutoff is superior to a static cutoff.

3. Increasing the maximum cutoff beyond some value does not increase the total accuracy further, because the error of the internal particles eventually dominates. At this point a larger minimum cutoff is required.

---

[10]This can be achieved by reducing the $\alpha$ value of Equation 8.

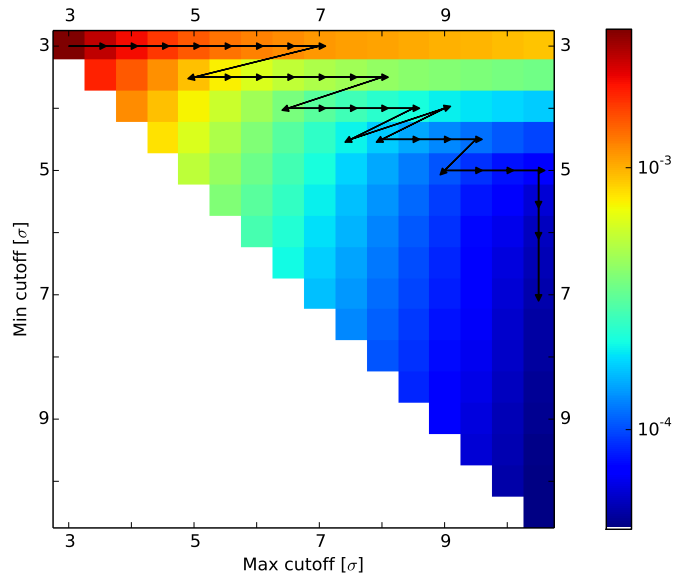

Figure 36: RMS error of the forces $\Delta F^{rel}$. The origin of each arrow denotes that this configuration is preferable w.r.t. performance while meeting a certain accuracy threshold. Arrow direction indicate the sequence in which to pick optimal parameters for an increasing accuracy threshold.

6.1.2. Density, Surface Tension and Energy

The surface tension, its uncertainty and the particle density have been computed in the same way as proposed in [15].

As it is evident from Figure 37, the DCM results for the density and surface tension converge to the results of PPPM as we increase the maximum cutoff. The monotonic convergence of the DCM results toward the PPPM solution once again shows the proper behavior of DCM. Figure 37 reveals many noteworthy properties: First, DCM with $r_c^{\min} = r_c^{\max}$ gives exactly the same results as the static $r_c$ version. Second, the results for DCM and the static cutoff are almost identical for cutoffs of $r_c \geq 8\sigma$, although DCM achieves these results $4-5$ times faster. Hence, we do not trade accuracy for performance. For instance, if one compares DCM 3.0/8.0 against static 5.0 much more accurate results are obtained in just a fraction of the time for the static 5.0 case. Third, the DCM results with $r_c^{max} \geq 8.0$ can be considered identical to those of the PPPM because the relative difference in the density between DCM and PPPM is less than 1% and because the slightly larger relative difference in the surface tension can be corrected by a tail correction (not shown). Moreover, PPPM is also an approximation to the correct solution and itself has an intrinsic error. Fourth, the surface tension and density monotonically increase as larger maximum cutoffs keep the particles closer together and preserve the interface more accurately.

---

[11]This is especially true as the ratio between interfacial and internal particles decreases.

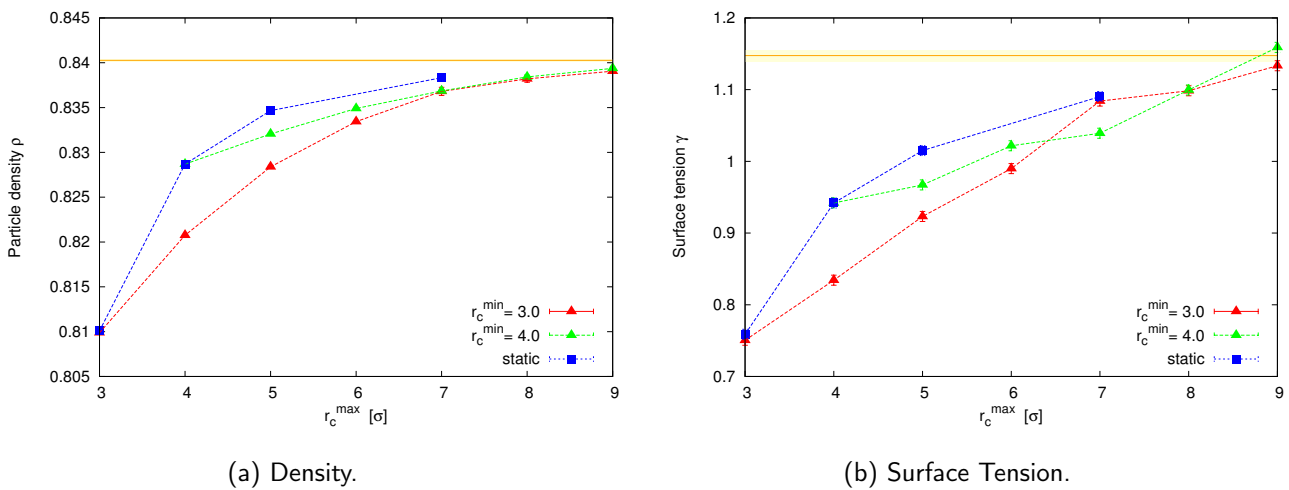

(a) Density.    (b) Surface Tension.

Figure 37: Solid line denotes the PPPM results.

Table 1 shows the energy of the DCM in comparison to the results of the static cutoff method and PPPM. Again, we see that DCM behaves as expected: as the maximum cutoff increases, we approach the PPPM and static solutions. Moreover, we also observe that the results of DCM $r_c^{min}/r_c^{max}$ are right in between static $r_c = r_c^{min}$ and static $r_c = r_c^{max}$, exactly as we would expect.

|  | Total energy | Std. deviation |
| --- | --- | --- |
| static 3.0 | -4.415 | 0.013 |
| static 5.0 | -4.772 | 0.016 |
| static 7.0 | -4.830 | 0.013 |
| DCM 3.0/5.0 | -4.557 | 0.013 |
| DCM 3.0/7.0 | -4.636 | 0.017 |
| DCM 3.0/8.0 | -4.660 | 0.016 |
| DCM 4.5/9.5 | -4.805 | 0.014 |
| PPPM | -4.859 | 0.013 |

Table 1: Total energy.

## 6.2. Performance

This subsection compares the performance of DCM against the statical cutoff method and PPPM. We use a Intel Xeon E5-2670 cluster for small test cases and the JUQUEEN supercomputer at Forschungszentrum Jülich for large scalability measurements. If not otherwise mentioned, we use 16 cores (i.e., two Intel Xeon E5-2670 CPUs). The minimum and maximum cutoff of DCM are respectively set to $r_c^{min} = 3.0$ and $r_c^{max} \geq 8.0$, such that the DCM results match the results of PPPM.

Figure 38 shows the speedup of DCM over its static counterpart. It shows the benefit of using the dynamic cutoff idea with respect to performance. The shown speedups range from $4.8\times$ to $5.7\times$. The speedup increases as the ratio between interfacial to internal particle decreases.

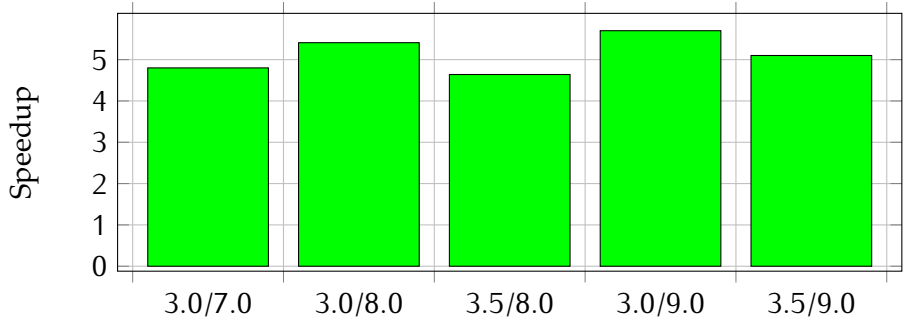

Figure 38: Speedup of DCM (using the exponential cutoff function) over a static cutoff with $r_c = r_c^{max}$ for a system with $192,000$ particles. Encoding: $r_c^{min}/r_c^{max}$.

### 6.2.1. MPI + OpenMP

Table 2 shows the speedup of our hybrid (MPI + OpenMP) DCM implementation over our pure MPI implementation. The total number of tasks (MPI ranks times threads) is kept constant for a varying number of threads. The results strengthen the intuition that a hybrid implementation is at least as good as a pure MPI implementation. In fact, the hybrid implementation increases the performance by up to 29%.

| #tasks | 2 threads | 4 threads | 8 threads | 16 threads |
|---|---|---|---|---|
| 1024 | 1.056 | 1.280 | 1.247 | 1.199 |
| 2048 | 1.084 | 1.121 | 1.291 | 1.229 |
| 4096 | 1.001 | 1.016 | 1.019 | 1.127 |

Table 2: Speedup of OpenMP + MPI over a pure MPI implementation. All simulations use the same system consisting of three million particles. Host: IBM BlueGene/Q.

### 6.2.2. Scalability

The scalability results of this section are conducted on the IBM BlueGene/Q system JUQUEEN at Forschungszentrum Jülich. JUQUEEN comprises $28,672$ nodes arranged in 28 racks. Each node consists of an IBM PowerPC A2 with 16 cores and 4-way simultaneous multi-threading (SMT). This yields a total of $458,752$ cores ($16,384$ cores per rack). Performance measurements indicate that SMT yields significant speedups (data not shown). Hence, all of our tests fully exploit the SMT feature of the IBM PowerPC running 4 tasks per core.

The domain for the weak scaling results is set up such that the volume $V^p = \frac{L_x}{P_x} \times \frac{L_y}{P_y} \times \frac{L_z}{P_z}$ and particle density per MPI process $p$ are kept constant with $L_x, L_y, L_z, P_x, P_y$ and $P_z$ being the computational domain and the number of MPI ranks in the $x, y$, and $z$ directions, respectively. The average number of particles per core is thus kept roughly constant at 1200 particles/core. Since LAMMPS applies domain decomposition we try to keep load balancing issues to a minimum by avoiding void spaces as much as possible. To be precise, $N = \#cores \cdot 1,200$ particles are randomly placed into a box $B$ of volume $L \times L \times 2L$. $B$ itself is then centered in a surrounding box of volume $L \times L \times (2L + d)$ with $d = 25.0\sigma$ such that the interface-interface distance $d$ (due to periodic boundary conditions) is sufficiently large. Strong scaling measurements use a system which is ideal for 32,768 cores (i.e., $32,768 \cdot 1,200 = 39,321,600$ particles).

Before we start our scalability discussion of DCM and PPPM, we stress that the settings for PPPM are chosen according to [15] and are considered to be optimal.

The red curve in Figure 39a shows the timings for DCM with sequential interface detection while the blue and purple curves show those timings with our parallel implementation of the interface detection. This shows both that parallel interface detection is clearly required and that our parallel implementation performs well. Moreover, DCM shows perfect weak-scaling and surpasses PPPM at around 1,000 to 2,000 cores. This makes DCM the preferred choice for large MD simulations on highly parallel systems.

The ideal strong scaling of DCM is shown in Figure 39b: while PPPM slows down significantly around 4000 to 8000 cores, DCM does not.

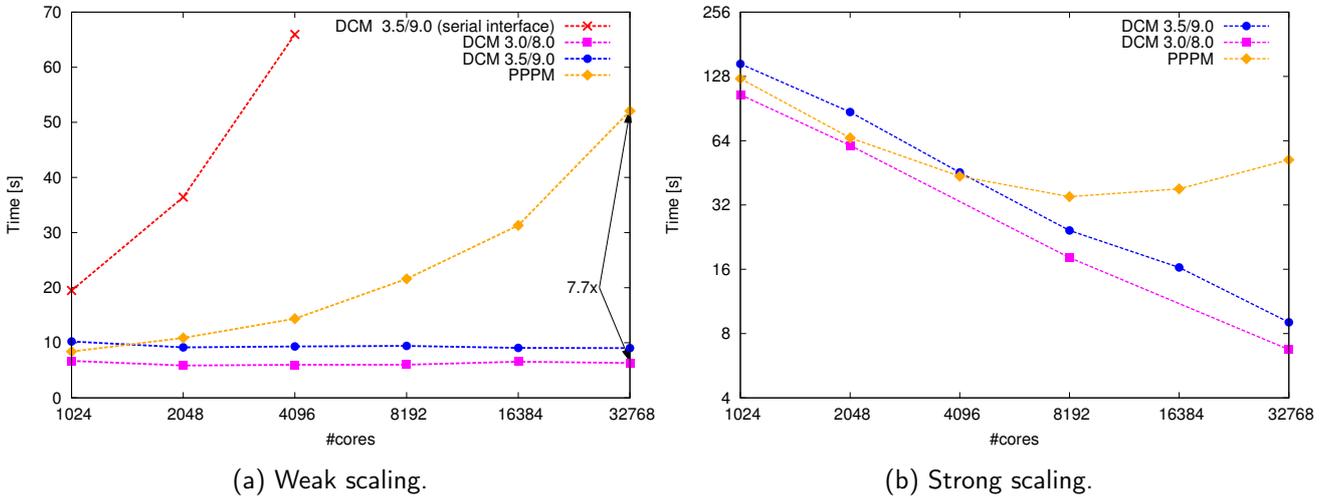

Figure 39: Scalability of DCM vs. PPPM. Results are accumulated over 200 MD steps.

Figure 40a shows the runtime of the neighbor-list build (NL) including interface detection (Interface), short-range Force calculation, communication (Comm) and remainder (Other) separately. Figure 40a highlights several interesting properties of the DCM. First, the communication time remains more or less constant, as we would expect from an algorithm with strictly local communications. In addition, the interface-detection contributes less than 1% of the overall runtime.

Figure 40b breaks the interface detection method down into its subroutines (see Section 4). We observe that $75-80\%$ of the interface detection time is due to volume segmentation. Moreover, the time it takes to segment the volume increases slightly when going from 1024 cores to 16384 cores. This is the source of a residual scaling issue that does not affect the overall scalability of DCM. The remaining subroutines behave nicely and have only a minor effect on the runtime of the interface detection routine.

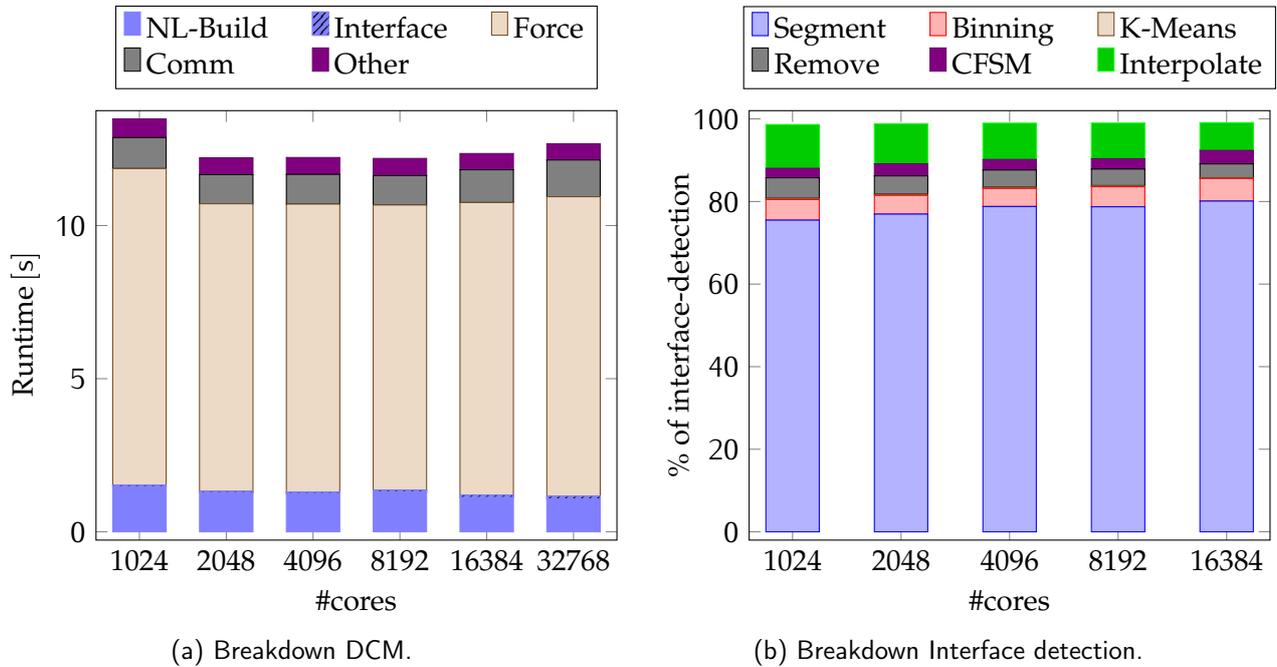

(a) Breakdown DCM.      (b) Breakdown Interface detection.

Figure 40: Breakdown of (a) DCM 4.5/9.5 and (b) its interface detection for a weak scaling simulation with 1200 particles per core. Results are accumulated over 200 MD steps.

## 7. Future Work

As we have mentioned several times throughout this thesis, there are a number of items left to be explored in the future. This section lists some possible future directions for the dynamic cutoff method with respect to performance, features, and accuracy.

**Performance.** Since DCM avoids the use of Newton's third law (i.e., has a good memory access pattern) and requires large cutoffs to work (i.e., it is a compute-bound algorithm), it is a perfect candidate for co-processors such as GPUs [36, 37] or Intel's Xeon Phi [50]. The work of Nguyen et al. [51] indicates that the speedup due to GPUs increases as the cutoff becomes larger. Another optimization along those lines would be to vectorize the two most important kernels —the force calculation and the neighbor-list build—via C/C++ AVX2 intrinsics. Pennycook et al. [50] show promising speedups from using explicit vectorization in MD simulations.

So far we have restricted ourselves to periodic boundary conditions. This, however, made it necessary to introduce a void space to separate the interfaces from each other. Hence, we want to extend DCM to allow non-PBCs as well and apply this idea to the same system that we have studied so far.

Moreover, a careful design of the cutoff function with respect to both accuracy and performance is left as future work.

**Accuracy.** Although the interface detection method reproduced the interface well for our purposes, we would like to carry out more accuracy studies and improve the accuracy of the detected interface if required.

**Features.** Moreover, this thesis applied the DCM only to a single-component Lennard-Jones system. The next step would be to extend DCM's capabilities to multi-component systems and apply the dynamic cutoff idea to other short-range potentials.

Although we have applied DCM only to a system with a planar liquid-vapor interface, we believe that DCM is also applicable to systems with non-planar and liquid-liquid interfaces as well.

As pointed out in Section 3.1 we want to extend the dynamic cutoff principle not only to particles within the liquid phase but also to those particles within the vapor phase as well.

A possibility to increase the accuracy of DCM quite substantially is to incorporate tail correction to the potential at runtime [34]. Given that we know the approximate volume $V^L$ and $V^V$ of the liquid and vapor phase and their approximate particle density $\rho^L$ and $\rho^V$, we could follow the ideas of Mecke et al. [34] and compute the force correction $\Delta F_i^{\text{tail}}$ for each particle $i$ according to:

$$\Delta F_i^{\text{tail}} = -\frac{d(\Delta u_i^{\text{tail}})}{dr} \quad (21)$$

where

$$\Delta u_i^{\text{tail}} = \int_{\substack{r_{ij} > r_c \\ r_j \in V^L}} u_{ij}(r_{ij}) \rho^L \, dr_{ij} + \int_{\substack{r_{ij} > r_c \\ r_j \in V^V}} u_{ij}(r_{ij}) \rho^V \, dr_{ij} \quad (22)$$

at every timestep. This correction does not require particle-particle interactions and is rather inexpensive to compute. Due to the rapid decay of short-range potentials, one could restrict the evaluation of the integral to the proximity of each particle, say $2r_c$, to keep the communication local.

## 8. Conclusion

We have developed a dynamic cutoff method for computing long-range dispersion interactions in molecular simulations which is based on computing the distance between particles and the interface. We also present a scalable, linear-time interface detection method for non-planar interfaces. This interface detection method enabled us to develop the DCM for short-range potentials. DCM is specially tailored for massively parallel supercomputers simulating interfacial systems with millions of particles.

We have implemented DCM as part of LAMMPS and showed that it exhibits desired properties such as (1) linear-time complexity, (2) local communication, and (3) ideal weak- and strong-scaling. Moreover, our accuracy results show that DCM is able to achieve the same accuracy as state-of-the-art algorithms for interfacial Lennard-Jones systems, while outperforming them for large systems. For instance, looking at a system with 39.3 million particles, DCM achieved the same accuracy as PPPM but was 7.7 times faster. While more case studies are required and more DCM features are desirable, our preliminary results indicate that DCM is a promising algorithm for massively parallel supercomputers.

We also introduced an interface detection method which is highly scalable and fast enough to be applicable in real time throughout the course of a MD simulation. As such, it might open the door to a wide variety of new MD applications.

## A. Parallel Fast Sweeping Method

We like to point out that the parallel implementation of the fast sweeping method by Zhao et al. [47] potentially results in many more floating-point operations than the serial implementation would require.

As an example, let $D \in \mathbb{R}^{n \times n}$ be the two dimensional domain of interest, $D_{i,j} \in \mathbb{R}^{n/2 \times n/2}$, $D_{i,j} \subset D$ are the subdomains of four processes $p_{i,j}$, $0 \leq i, j < 2$. Let $N = n \times n$ be the total number of grid points. Imagine that we place a single point into the center of $D_{0,0}$. All processes would start to process their subdomain requiring $\approx N/4$ operations each (i.e., $N$ operations in total). However, all processes except for $p_{0,0}$ would do useless work. At the end of the first computational step only $D_{0,0}$ has been computed correctly. The next step is to exchange the solution at the boundary of each subdomain with its neighbors. Once the communication is done, each process again starts to process its subdomain again requiring $\approx N/4$ operations each. At the end of this second computation phase $D_{0,1}$ $D_{1,0}$ have also been computed correctly. However, to compute $D_{1,1}$ correctly another computation step is required. All in all this algorithm took roughly three times more operations than the serial implementation and as a consequence results in poor scaling.